\documentclass[journal]{IEEEtran}
\usepackage{eurosym}
\usepackage{cite,algorithm,algpseudocode,array,url}
\usepackage{graphicx,multirow,colortbl}
\usepackage[percent]{overpic}
\usepackage{amsfonts,times,amsmath,amssymb,amsthm}
\usepackage[math]{cellspace}

\ifCLASSOPTIONcompsoc
\else
\fi
\ifCLASSOPTIONcaptionsoff
\let\MYoriglatexcaption\caption
 \renewcommand{\caption}[2][\relax]{\MYoriglatexcaption[#2]{#2}}
\fi
\newcounter{MYtempeqncnt}
\theoremstyle{plain}
\newtheorem{theorem}{Theorem}

\newtheorem{proposition}[theorem]{Proposition}

\theoremstyle{definition}
\newtheorem{definition}{Definition}

\theoremstyle{remark}
\newtheorem*{remark}{Remark}

\begin{document}

\title{A Generalized Labeled Multi-Bernoulli Filter Implementation using
Gibbs Sampling}
\author{Hung~Gia~Hoang, Ba-Tuong~Vo, Ba-Ngu~Vo
\thanks{Acknowledgement: This work is supported by the Australian Research Council
under the Future Fellowship FT0991854 and Discovery Early Career Researcher
Award DE120102388.}
\thanks{
Part of this paper is presented at the 18th Fusion conference, Washington, DC, USA, July 6-9, 2015.}
\thanks{H.~G. Hoang, B.-N. Vo, and B.-T. Vo are with the Department of Electrical
and Computer Engineering, Curtin University, Bentley, WA 6102, Australia
(email: \{hung.hoang,ba-ngu.vo,ba-tuong.vo\}@curtin.edu.au).}}
\maketitle

\begin{abstract}
This paper proposes an efficient implementation of the generalized labeled
multi-Bernoulli (GLMB) filter by combining the prediction and update into a
single step. In contrast to the original approach which involves separate
truncations in the prediction and update steps, the proposed implementation
requires only one single truncation for each iteration, which can be
performed using a standard ranked optimal assignment algorithm. Furthermore,
we propose a new truncation technique based on Markov Chain Monte Carlo
methods such as Gibbs sampling, which drastically reduces the complexity of
the filter. The superior performance of the proposed approach is
demonstrated through extensive numerical studies.
\end{abstract}





\section{Introduction}

Multi-object tracking refers to the problem of jointly estimating the number
of objects and their trajectories from sensor data. Driven by aerospace
applications in the 1960's, today multi-object tracking lies at the heart of
a diverse range of application areas, see for example the texts \cite%
{BSF88,BP99, Mah07, MallickBook12, mahler2014advances}. The most
popular approaches to multi-object tracking are the joint probabilistic data
association filter \cite{BSF88}, multiple hypothesis tracking \cite%
{BP99}, and more recently, random finite set (RFS) \cite{Mah07,
mahler2014advances}.

The RFS approach has attracted significant attention as a general systematic
treatment of multi-object systems and provides the foundation for the
development of novel filters such as the Probability Hypothesis Density
(PHD) filter \cite{MahlerPHD,VSD05, VoMaGMPHD05}, Cardinalized PHD (CPHD)
filter \cite{MahlerCPHD,VoGaussianCPHD07}, and multi-Bernoulli filters \cite%
{Mah07,VVC09,VVPS10}. While these filters were not designed to estimate the
trajectories of objects, they have been successfully deployed in many
applications including radar/sonar \cite{TobiasLanterman05}, \cite%
{TobiasLanterman08}, \cite{ClarkBell05}, computer vision \cite%
{MTC_CSVT08,HVVS_PR12,HVV_TSP13}, cell biology \cite{Rez_TMI15}, autonomous
vehicle \cite{MVA_TRO11, LHG_TSP11, LCS_SLAM_STSP13, AVMM_SLAM14},
automotive safety \cite{Bat08,MRD13}, sensor scheduling \cite%
{RV10,RVC11,HV14sencon, Gostar13} and sensor network \cite%
{Zhang_TAC11,BCF_STSP13,UCJ_STSP13}.

The introduction of the generalized labeled multi-Bernoulli (GLMB) RFS in 
\cite{VoGLMB13} has led to the development of the first tractable RFS-based
multi-object tracker - the $\delta $-GLMB filter. The $\delta$-GLMB filter
is attractive in that it exploits the conjugacy of the GLMB family to
propagate forward in time the (labeled) multi-object filtering density
exactly \cite{VoGLMB13}. Each iteration of this filter involves an update
operation and a prediction operation, both of which result in weighted sums
of multi-target exponentials with intractably large number of terms. The
first implementation of the $\delta $-GLMB filter truncate these sums by
using the $K$-shortest path and ranked assignment algorithms, respectively,
in the prediction and update to determine the most significant components 
\cite{VVP_GLMB13}. 

While the original two-staged implementation is intuitive and highly
parallelizable, it is structurally inefficient as it requires many
intermediate truncations of the $\delta $-GLMB densities. Specifically, in
the update, truncation is performed by solving a ranked assignment problem
for each predicted $\delta $-GLMB component. Since truncation of the
predicted $\delta $-GLMB sum is performed separately from the update, in
general, a significant portion of the predicted components would generate
updated components with negligible weights. Hence, computations are wasted
in solving a large number of ranked assignment problems, each of which has
cubic complexity in the number of measurements.

In this paper, we present a new implementation by formulating a joint
prediction and update that eliminates inefficient truncation
procedures in the original approach. The key innovation is the exploitation
of the direct relationship between the components of the $\delta $-GLMB
filtering densities at consecutive iterations to circumvent solving a ranked
assignment problem for each predicted component. In contrast to the original
implementation, the proposed joint implementation only requires one
truncation per component in the filtering density.

Naturally, the joint prediction and update allows truncation of the $\delta $%
-GLMB filtering density (without explicitly enumerating all the components)
using the ranked assignment algorithm \cite{Murty68}, \cite{Pascoaletal03}, 
\cite{Pedersenetal08}. More importantly, it admits a very efficient
approximation of the $\delta $-GLMB filtering density based on Markov Chain
Monte Carlo methods. The key innovation is the use of Gibbs sampling to
generate significant updated $\delta $-GLMB components, instead of
deterministically generating them in order of non-increasing weights. The
advantages of the proposed stochastic solution compared to the rank
assignment algorithm are two-fold. First, it eliminates unnecessary
computations incurred by sorting the\ components, and reduces the complexity
from cubic to linear in the number of measurements. Second, it automatically
adjusts the number of significant components generated by exploiting the
statistical characteristics of the component weights. 

The paper is organized as follows. Background on labeled RFS and the $\delta 
$-GLMB filter is provided in section \ref{sec:BG}. Section \ref%
{sec:fast_impl} presents the joint prediction and update formulation and the
Gibbs sampler based implementation of the $\delta $-GLMB filter. Numerical
results are presented in Section \ref{sec:sim} and concluding remarks are
given in Section \ref{sec:sum}.

\section{Background}\label{sec:BG} 
This section summarizes the labeled RFS and the GLMB filter
implementation. We refer the reader to the original work \cite{VoGLMB13,
VVP_GLMB13} for detailed expositions.

For the rest of the paper, single-object states are represented by lowercase
letters, e.g. $x$, $\mathbf{x}$ while multi-object states are represented by
uppercase letters, e.g. $X$, $\mathbf{X}$, symbols for labeled states and
their distributions are bolded to distinguish them from unlabeled ones, e.g. 
$\mathbf{x}$, $\mathbf{X}$, $\mathbf{\pi }$, etc, spaces are represented by
blackboard bold e.g. $\mathbb{X}$, $\mathbb{Z}$, $\mathbb{L}$, $\mathbb{N}$,
etc, and the class of finite subsets of a space $\mathbb{X}$ is denoted by $%
\mathbf{\mathcal{F}(}\mathbb{X)}$. We use the standard inner product
notation $\left\langle f,g\right\rangle \triangleq \int f(x)g(x)dx,$ and the
following multi-object exponential notation $h^{X}\triangleq
\prod\nolimits_{x\in X}h(x)$, where $h$ is a real-valued function, with $%
h^{\emptyset }=1$ by convention. We denote a generalization of the Kronecker
delta that takes arbitrary arguments such as sets, vectors, etc, by 
\begin{equation*}
\delta _{Y}(X)\triangleq \left\{ 
\begin{array}{l}
1,\text{ if }X=Y \\ 
0,\text{ otherwise}%
\end{array}%
\right. ,
\end{equation*}%
and the inclusion function, a generalization of the indicator function, by%
\begin{equation*}
1_{Y}(X)\triangleq \left\{ 
\begin{array}{l}
1,\text{ if }X\subseteq Y \\ 
0,\text{ otherwise}%
\end{array}%
\right. .
\end{equation*}%
We also write $1_{Y}(x)$ in place of $1_{Y}(\{x\})$ when $X$ = $\{x\}$.

\subsection{Labeled RFS}

\label{subsec:LabeledRFS} A labeled RFS is simply a finite set-valued random
variable where each single-object dynamical state is augmented with a unique
label that can be stated concisely as follows

\begin{definition}
A labeled RFS with state space $\mathbb{X}$ and (discrete) label space $%
\mathbb{L}$ is an RFS on $\mathbb{X}\mathcal{\times }\mathbb{L}$ such that
each realization has distinct labels.
\end{definition}

Let $\mathcal{L}:\mathbb{X}\mathcal{\times }\mathbb{L}\rightarrow \mathbb{L}$
be the projection $\mathcal{L}((x,\ell ))=\ell $, then a finite subset set $%
\mathbf{X}$ of $\mathbb{X}\mathcal{\times }\mathbb{L}$ has distinct labels
if and only if $\mathbf{X}$ and its labels $\mathcal{L}(\mathbf{X})=\{%
\mathcal{L}(\mathbf{x})\!:\!\mathbf{x}\!\in \!\mathbf{X}\}$ have the same
cardinality, i.e. $\delta _{|\mathbf{X}|}(|\mathcal{L(}\mathbf{X})|)=1$. The
function $\Delta (\mathbf{X})\triangleq $ $\delta _{|\mathbf{X}|}(|\mathcal{%
L(}\mathbf{X})|)$ is called the \emph{distinct label indicator}.

The set integral defined for any function $\mathbf{f:\mathcal{F}(}\mathbb{X}%
\mathcal{\times }\mathbb{L)}\rightarrow \mathbb{R}$ is given by%
\begin{equation*}
\int \mathbf{f}(\mathbf{X})\delta \mathbf{X}=\sum_{i=0}^{\infty }\frac{1}{i!}%
\int \mathbf{f}(\{\mathbf{x}_{1},...,\mathbf{x}_{i}\})d(\mathbf{x}_{1},...,%
\mathbf{x}_{i}).
\end{equation*}
where the integral of a function $\mathbf{f:}\mathbb{X}\mathcal{\times }%
\mathbb{L}\rightarrow \mathbb{R}$ is: 
\begin{equation*}
\int \mathbf{f}(\mathbf{x})d\mathbf{x}=\sum_{\ell \in \mathbb{L}}\int_{%
\mathbb{X}}\mathbf{f}((x,\ell ))dx,
\end{equation*}

The notion of labeled RFS enables the incorporation of individual object
identity into multi-object system and the Bayes filter to be used as a
tracker of these multi-object states.

\subsection{Bayes filter for labeled RFS}

\label{subsec:RFSBayes} Suppose that at time $k$, there are $N_{k}$ target
states $\mathbf{x}_{k,1},\ldots ,\mathbf{x}_{k,N_k}$, each taking values in
the (labeled) state space $\mathbb{X\times L}$. In the random finite set
formulation the set of targets is treated as the \emph{multi-object state} 
\begin{equation}
\mathbf{X}_{k} =\{\mathbf{x}_{k,1},\ldots ,\mathbf{x}_{k,N_k}\}.
\end{equation}
Each state $(x_{k},\ell) \in \mathbf{X}_{k}$ either survives with
probability $p_S(x_{k},\ell)$ and evolves to a new state $(x_{k+1},\ell)$ or
dies with probability $1-p_S(x_{k},\ell)$. The dynamics of the survived
targets are encapsulated in the multi-object transition density $\mathbf{f}_{k+1|k}(\mathbf{X}_{}|\mathbf{X}_{k}$).

For a given multi-object state $\mathbf{X}_k$, each state $%
(x_k,\ell) \in \mathbf{X}_{k}$ at time $k$ is either detected with probability $%
p_D(x_k,\ell)$ and generates an observation $z$ with likelihood $%
g(z|x_k,\ell)$ or missed with probability $1-p_D(x_k,\ell)$. The \emph{multi-object observation} at time $k$, $Z_{k} =\{z_{k,1},\ldots ,z_{k,M_k}\}$%
, is the superposition of the observations from detected states and Poisson
clutters with intensity $\kappa$. Assuming that, conditional on $\mathbf{X}%
_{k}$, detections are independent, and that clutter is independent of the
detections and is distributed as a Poisson RFS, the multi-object likelihood is given by \cite%
{VoGLMB13,VVP_GLMB13} 
\begin{equation}
g(Z_k|\mathbf{X}_k)=e^{-\left\langle \kappa ,1\right\rangle }\kappa
^{Z_k}\sum_{\theta^{_{\!}} \in \Theta_{_{\!}\mathcal{L}_{_{\!}}(_{_{\!}}\mathbf{X}_{k_{\!}})}}\left[ \psi
_{\!Z_{_{\!}k\!}}(_{\!}\cdot ;\theta_{\!})\right]^{\mathbf{X}_k}  \label{eq:RFSmeaslikelihood0}
\end{equation}%
where $\theta_{\!}: \mathbb{L} \to \{0,1,\ldots,|Z_k|\}$ is a function such that $\theta_{\!}{(i)}=\theta_{\!}{(i^\prime)} > 0$ implies $i=i^\prime$, and 
\begin{equation}
\psi _{Z_{_{\!}k\!}}(x,\ell ;\theta_{\!})=\left\{ 
\begin{array}{ll}
\frac{p_{D}(x,\ell )g(_{\!}z_{\theta_{\!}(_{\!}\ell_{\!})\!}|_{_{\!}}x,\ell )}{\kappa(_{\!}z_{\theta_{\!}(_{\!}\ell_{\!})\!})}, & \text{if }\theta_{_{\!}}(\ell )>0 \\ 
1-p_{D}(x,\ell ), & \text{if }\theta_{_{\!}}(\ell )=0%
\end{array}%
\right.  \label{eq:PropConj5}
\end{equation}

\begin{remark}
$\theta_{\!}$ is called an \emph{association map} since it provides the mapping
between tracks and observations, i.e. which track generates which
observation, with undetected tracks assigned to 0. The condition $\theta_{\!}{(i)}=\theta_{\!}{(i^\prime)} > 0$ implies $i=i^\prime$ ensures that a track can generate at most one measurement at a point of time.

The image of a set $I \subset \mathbb{L}$ through the map $\theta_{}$ is denoted by $\theta_{\!}(I)$, i.e.
\[\theta_{\!}(I) =\{\theta_{\!}(_{\!}\ell_{\!}): \ell \in I\},\]
while the notation $\Theta_{_{\!}I}$ is used to denote the collection of all eligible association maps on domain $I$, i.e.
\[\Theta_{_{\!}I} = \left\{\theta_{\!}: \theta^{-1}\!\left(\{0,1,\ldots,|Z_k|\}\right)=I\right\}\]
\end{remark}

If the clutter is distributed as an iid cluster RFS, i.e. $\pi_C(K)=\rho(|K|)|K|![p_C(\cdot)]^K$, the multi-object likelihood is
\begin{equation}
g(Z_k|\mathbf{X}_k)\!=\!\!\!\sum_{Z_{\bar{C}}\subseteq Z_k}\!\!\!\pi_C(Z_k\!\!\setminus\!\!Z_{\bar{C}})\!\!\!\sum_{\theta_{\bar{C}}\in\Theta_{\mathcal{L}(\mathbf{X}_k)}}\!\!\!g_M(Z_{\bar{C}}|\mathbf{X}_k;\theta_{\bar{C}}),
\end{equation}
where $\theta_{\bar{C}}$ is the association map from $\mathcal{L}(\mathbf{X}_k)$ to $\{0,\ldots,|Z_{\bar{C}}|\}$ and 

Given a multi-object system as described above, the objective is to find the 
\emph{multi-object filtering density}, denoted by $\mathbf{\pi }_{k+1}(\mathbf{X}_{}
|Z_{k+1})$\footnote{For convenience, we drop the dependence on past measurements upto time $k$},
which captures all information on the number of targets and individual
target states at time $k+1$. In multi-object Baysian filtering, the
multi-object filtering density is computed recursively in time according to
the following prediction and update, commonly referred to as \emph{multi-object Bayes recursion} \cite{Mah07} 
\begin{align}
\!\!\mathbf{\pi }_{k+1|k\!}(\mathbf{X}_{}|Z_{k\!})& =\!\int \!\mathbf{f}_{k+1|k\!}(\mathbf{X}_{}|\mathbf{X}_{k\!})\mathbf{\pi}_{k\!}(\mathbf{X}
_{k}|Z_{k\!})\delta_{\!}\mathbf{X}_{k},  \label{eq:MTBayesPred} \\
\!\!\mathbf{\pi }_{k+1\!}(\mathbf{X}_{}|Z_{k+1\!})& =\!\frac{g(_{\!}Z_{k+1}|\mathbf{X}_{\!})\mathbf{\pi }_{k+1|k\!}(\mathbf{X}_{}|Z_{k\!})}{\int\! g(_{\!}Z_{k+1}|\mathbf{X}_{\!})\mathbf{\pi}_{k+1|k\!}(\mathbf{X}_{}|Z_{k\!})\delta_{\!}\mathbf{X}_{}}.  \label{eq:MTBayesUpdate}
\end{align}

Note, however, that the Bayes filter is intractable since the set integrals in \eqref{eq:MTBayesPred}-\eqref{eq:MTBayesUpdate} have no analytic solution in general.

\subsection{Delta generalized labeled multi-Bernoulli RFS}

\label{subsec:GLMB_RFS}

The $\delta$-GLMB RFS, a special class of labeled RFS, provides an exact
solution to \eqref{eq:MTBayesPred}-\eqref{eq:MTBayesUpdate}. This is because
the $\delta$-GLMB RFS is closed under the multi-object Chapman-Kolmogorov
equation with respect to the multi-object transition kernel and is conjugate
with respect to the multi-object likelihood function \cite{VoGLMB13}.

\begin{definition}
A $\delta$-GLMB RFS is a labeled RFS with state space $\mathbb{X}$ and
(discrete) label space $\mathbb{L}$, distributed according to 
\begin{equation}  \label{eq:generativeGLMB}
\mathbf{\pi }(\mathbf{X})=\Delta (\mathbf{X})\sum_{(I,\xi )\in \mathcal{F}(%
\mathbb{L})\times \Xi } \omega ^{(I,\xi )}\delta _{I}(\mathcal{L(}\mathbf{X}%
))\left[ p^{(\xi )}\right] ^{\mathbf{X}},
\end{equation}%
where $\Xi$ is a discrete space while $\omega ^{(I,\xi )}$ and $p^{(\xi )}$
satisfy 
\begin{align}
\sum_{(I,\xi )\in \mathcal{F}(\mathbb{L})\times \Xi }\omega ^{(I,\xi )} &=1,
\\
\int p^{(\xi )}(x,\ell)dx&=1.
\end{align}
\end{definition}

\begin{remark}
The $\delta$-GLMB density is essentially a mixture of multi-object
exponentials, in which each components is identified by a pair $(I,\xi)$.
Each $I \in \mathcal{F}(\mathbb{L})$ is a set of tracks labels while $\xi
\in \Xi$ represents a history of association maps $\xi=(\theta_1,\ldots,%
\theta_k)$. The pair $(I,\xi)$ can be interpreted as the hypothesis that the
set of tracks $I$ has a history of $\xi$ association maps and corresponding
kinematic state densities $p^{\!(\xi)}$. The weight $\omega ^{(I,\xi
)}\delta _{I}(\mathcal{L(}\mathbf{X}))$, therefore, can be considered as the
probability of the hypothesis $(I,\xi )$.
\end{remark}

Denote the collection of all label sets with $n$ unique elements by $\mathcal{F}_n(\mathbb{L})$, the cardinality distribution of a $\delta$-GLMB
RFS is given by 
\begin{equation}  \label{card_dis}
\mathbf{\rho}(n) = \sum_{(I,\xi )\in \mathcal{F}_n(\mathbb{L})\times \Xi
}\omega ^{(I,\xi )}.
\end{equation}

\begin{figure}[htb]
\centering
\includegraphics[scale=.8]{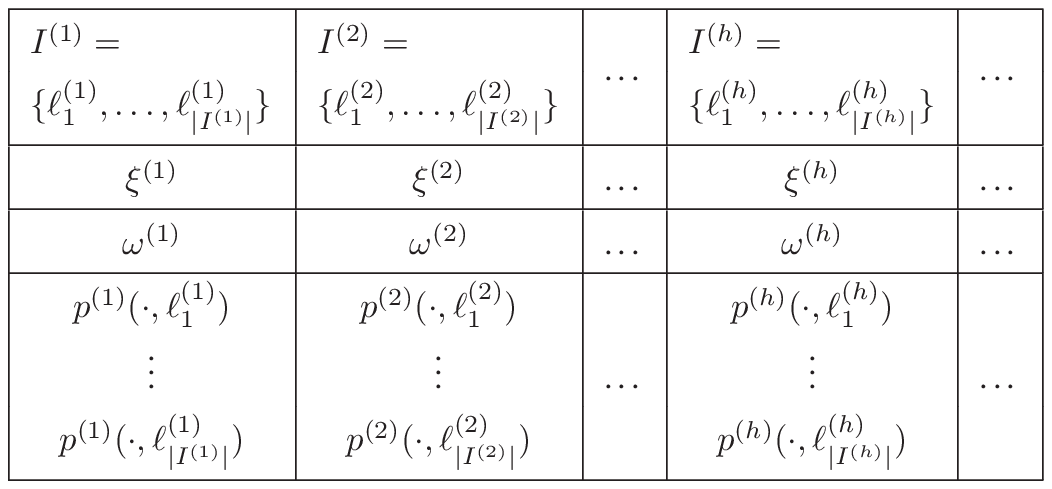}
\caption{An enumeration of a $\protect\delta $-GLMB parameter set with each
component indexed by an integer $h$. The hypothesis\ for component $h$ is $%
(I^{(h)},\protect\xi ^{(h)})$ while its weight and associated track
densities are $\protect\omega ^{(h)}$ and $p^{(h)}(\cdot ,\ell )$, $\ell \in
I^{(h)}$.}
\label{fig:paramtable}
\end{figure}

A $\delta $-GLMB is completely characterized by the set of parameters $\{(\omega ^{(I,\xi )},p^{(\xi )}):(I,\xi )\in \mathcal{F}\!(\mathbb{L})\!\times \!\Xi \}$. For implementation it is convenient to consider the set of parameters as an enumeration of all $\delta $-GLMB components (with
positive weight) together with their associated weights and track densities $\{(I^{(h)},\xi ^{(h)},\omega ^{(h)},p^{(h)})\}_{h=1}^{H}$, as shown in
Figure~\ref{fig:paramtable}, where $\omega ^{(h)}\triangleq \omega^{(I^{(h)},\xi^{(h)})}$ and $p^{(h)}\triangleq p^{(\xi ^{(h)})}$. 

Given a $\delta$-GLMB initial density, all subsequent multi-object densities are $\delta$-GLMBs and can be computed exactly by a tractable filter called the $\delta $-GLMB filter.

\subsection{Delta generalized labeled multi-Bernoulli filter}
\label{subsec:GLMB_filt}

The $\delta $-GLMB filter recursively propagates a $%
\delta $-GLMB density forward in time via the Bayes recursion equations \eqref{eq:MTBayesPred} and \eqref{eq:MTBayesUpdate}. Closed
form solutions to the prediction and update of the $\delta $-GLMB filter are
given in the following propositions \cite{VoGLMB13}.

\begin{proposition}
\label{Prop_CK_strong} If the multi-target posterior at time $k-1$ is a $\delta$-GLMB of the form \eqref{eq:generativeGLMB}, i.e. 
\begin{eqnarray}\label{eq:priorGLMB}
\mathbf{\pi}_{_{\!}k_{\!}-_{\!}1{\!}}(_{_{\!}}\mathbf{X}_{_{\!}}|_{_{\!}}Z_{_{\!}k_{\!}-_{\!}1{\!}})\!\!\!&=&\!\!\!\Delta_{\!}(_{_{\!}}\mathbf{X}_{_{\!}})\!\!\!\!\!\!\sum_{\!(\!I_{_{\!}k_{\!}-_{\!}1{\!}},\xi_{_{\!}k_{\!}-_{\!}1_{\!}})\in\mathcal{F}_{_{\!}}(_{_{\!}}\mathbb{L}_{_{\!}})_{\!}\times_{\!}\Xi}\!\!\!\!\!\!\omega_{_{\!}k_{\!}-_{\!}1{\!}}^{(\!I_{_{\!}k_{\!}-_{\!}1_{\!}},\xi_{_{\!}k_{\!}-_{\!}1{\!}})}\times\notag\\
&&\delta_{_{\!}I_{_{\!}k_{\!}-_{\!}1}\!}(_{\!}\mathcal{L}(_{\!}\mathbf{X}_{\!})_{\!})\!\left[_{\!}p_{_{\!}k_{\!}-_{\!}1}^{_{\!}(_{\!}\xi_{_{\!}k_{\!}-_{\!}1\!})\!}(_{\!}\cdot_{_{\!}}|_{\!}Z_{_{\!}k_{\!}-_{\!}1\!})_{\!}\right]^{\!\mathbf{X}_{}}\!\!\!,
\end{eqnarray}
and the birth density $\mathbf{f}_B$ is defined on $\mathcal{F}(\mathbb{X} \times 
\mathbb{B})$ according to 
\begin{equation}  \label{eq:birthGLMB}
\mathbf{f}_{_{\!}B_{_{\!}}}(_{_{\!}}\mathbf{Y}_{_{\!}})=\Delta_{\!}(_{_{\!}}\mathbf{Y}_{_{\!}})\omega_{_{\!}B_{\!}}(_{\!}\mathcal{L}_{_{\!}}(_{_{\!}}\mathbf{Y}_{\!})_{\!})[p_{_{\!}B_{\!}}(_{\!}\cdot_{\!})]^{\!\mathbf{Y}},
\end{equation}
then the multi-target prediction density to the next time is a $\delta $-GLMB given
by 
\begin{eqnarray}  \label{eq:PropCKstrong1}
\!\!\!\!\!\!\!\!\!\!\mathbf{\pi}_{_{\!}k_{_{\!}}|_{_{\!}}k_{\!}-_{\!}1{\!}}(_{_{\!}}\mathbf{X}_{_{\!}}|_{_{\!}}Z_{_{\!}k_{\!}-_{\!}1\!})
\!\!\!\!&=&\!\!\!\! \Delta_{{\!}}(_{_{\!}}\mathbf{X}_{_{\!}})\times  \notag \\
&&\!\!\!\!\!\!\!\!\!\!\!\!\!\!\!\!\!\!\!\!\sum_{(\!I_{_{\!}k_{\!}},\xi_{_{\!}k_{\!}-_{\!}1{\!}})%
\in\mathcal{F}_{_{\!}}(_{\!}\mathbb{L}\cup_{}\mathbb{B}_{\!})_{\!}\times_{\!}\Xi}\!\!\!\!\!\!\!\!\!\!\!
\omega_{_{\!}k_{_{\!}}|_{_{\!}}k_{\!}-_{\!}1{\!}}^{(\!I_{_{\!}k_{\!}},\xi_{_{\!}k_{\!}-_{\!}1\!})\!}\delta_{_{\!}I_{_{\!}k}\!}(_{\!}\mathcal{L}_{\!}(_{\!}\mathbf{X}_{\!})_{\!})\!\left[_{\!}p_{_{\!}k_{_{\!}}|_{_{\!}}k_{\!}-_{\!}1{\!}}^{_{\!}(_{\!}\xi_{_{\!}k_{\!}-_{\!}1\!})\!}(_{\!}\cdot_{_{\!}}|_{\!}Z_{_{\!}k_{\!}-_{\!}1{\!}})_{\!}\right]^{\!\mathbf{X}_{}}\!\!\!,
\end{eqnarray}
where\allowdisplaybreaks%
\begin{eqnarray}
\!\!\!\!\!\omega_{_{\!}k_{_{\!}}|_{_{\!}}k_{\!}-_{\!}1{\!}}^{(\!I_{_{\!}k_{\!}},\xi_{_{\!}k_{\!}-_{\!}1\!})}\!\!\!\!\!&=&\!\!\!\!\!\omega_{_{\!}k_{\!}-_{\!}1{\!}}^{(\!I_{_{\!}k_{\!}-_{\!}1_{\!}},\xi_{_{\!}k_{\!}-_{\!}1{\!}})}\omega_{_{\!}S}^{_{\!}(_{\!}\xi_{_{\!}k_{\!}-_{\!}1\!})\!}(_{\!}I_{_{\!}k\!}\cap_{\!} \mathbb{L}_{\!})\omega_{_{\!}B_{\!}}(_{\!}I_{_{\!}k\!}\cap_{\!} \mathbb{B}_{\!})  \label{eq:PropCKstrongwp} \\
\!\!\!\!\!\omega_{_{\!}S}^{_{\!}(_{\!}\xi_{_{\!}k_{\!}-_{\!}1\!})\!}(_{\!}L_{\!})\!\!\!\!\!&=&\!\!\!\!\!\left[_{\!}\eta_{_{\!}S\!}^{_{\!}(_{\!}\xi_{_{\!}k_{\!}-_{\!}1\!})\!}\right]^{\!L}\!\!\!\sum_{I_{_{\!}k_{\!}-_{\!}1_{\!}}\supseteq_{_{\!}}L}\!\!\left[_{\!}1\!-\!\eta_{_{\!}S\!}^{_{\!}(_{\!}\xi_{_{\!}k_{\!}-_{\!}1\!})\!}\right]^{\!I_{_{\!}k_{\!}-_{\!}1\!}-L}\label{eq:PropCKstrongws} \\
\!\!\!\!\!\eta_{_{\!}S\!}^{_{\!}(_{\!}\xi_{_{\!}k_{\!}-_{\!}1\!})\!}(_{\!}\ell)\!\!\!\!\!&=&\!\!\!\!\!\left\langle p_{_{\!}S}(\cdot ,\ell
),p_{_{\!}k_{\!}-_{\!}1}^{_{\!}(_{\!}\xi_{_{\!}k_{\!}-_{\!}1\!})\!}(_{\!}\cdot ,\ell)\right\rangle  \label{eq:PropCKstrong_eta} \\
\!\!\!\!\!p_{_{\!}k_{_{\!}}|_{_{\!}}k_{\!}-_{\!}1_{\!}}^{_{\!}(_{\!}\xi_{_{\!}k_{\!}-_{\!}1\!})_{\!}}(_{\!}x,\ell_{_{\!}})\!\!\!\!\!&=&\!\!\!\!\!1_{\mathbb{L}_{\!}}(_{\!}\ell_{_{\!}})p_{_{\!}S}^{_{\!}(_{\!}\xi_{_{\!}k_{\!}-_{\!}1\!})_{\!}}(_{\!}x,\ell_{_{\!}})+1_{_{\!}\mathbb{B}_{\!}}(_{\!}\ell_{_{\!}})p_{_{\!}B_{_{\!}}}(_{\!}x,\ell_{_{\!}})\label{eq:PropCKstrongpp} \\
\!\!\!\!\!p_{_{\!}S}^{_{\!}(_{\!}\xi_{_{\!}k_{\!}-_{\!}1\!})_{\!}}(_{\!}x,\ell_{_{\!}})\!\!\!\!\!&=&\!\!\!\!\!\frac{\left\langle p_{_{\!}S}(\cdot
,\ell)f_{_{\!}k_{_{\!}}|_{_{\!}}k_{\!}-_{\!}1_{\!}}(x|\cdot ,\ell ),p_{_{\!}k_{\!}-_{\!}1}^{_{\!}(_{\!}\xi_{_{\!}k_{\!}-_{\!}1\!})\!}(\cdot ,\ell )\right\rangle}{\eta_{_{\!}S\!}^{_{\!}(_{\!}\xi_{_{\!}k_{\!}-_{\!}1\!})\!}(\ell )}  \label{eq:PropCKstrongps}
\end{eqnarray}
\end{proposition}

\begin{proposition}
\label{PropBayes_strong} Given the prediction density in %
\eqref{eq:PropCKstrong1}, the multi-target posterior is a $\delta $-GLMB
given by%
\begin{eqnarray}
\!\!\!\!\!\!\!\!\!\mathbf{\pi}_{_{\!}k_{\!}}(_{_{\!}}\mathbf{X}_{_{\!}}|_{_{\!}}Z_{_{\!}k\!})\!\!\!\!&=&\!\!\!\!\Delta_{\!}(_{\!}\mathbf{X}_{\!}) \!\!\!\!\!\sum_{(\!I_{_{\!}k_{\!}},\xi_{_{\!}k_{\!}-_{\!}1\!},\theta_{_{\!}k_{\!}})\in\mathcal{F}\!(\mathbb{L} \cup \mathbb{B})\times \Xi \times \Theta_{_{\!}I_{_{\!}k}}}\!\!\!\!\!\!\!\omega_{_{\!}k}^{_{\!}(I_{_{\!}k_{\!}},\xi_{_{\!}k_{\!}-_{\!}1\!},\theta_{_{\!}k\!})\!}(_{\!}Z_{_{\!}k\!})\times  \notag \\
&&\!\!\!\!\delta_{I_{_{\!}k_{\!}}}(_{\!}\mathcal{L_{\!}(_{\!}}\mathbf{X}_{\!})_{\!})\!\left[p_{k}^{_{\!}(_{\!}\xi_{_{\!}k_{\!}-_{\!}1\!},\theta_{_{\!}k\!})}(\cdot|Z_{\!k}\!)\right] ^{\!\mathbf{X}_{}}\!\!\!,  \label{eq:PropBayes_strong0}
\end{eqnarray}
where \allowdisplaybreaks%
\begin{eqnarray}
\omega_{_{\!}k}^{_{\!}(_{\!}I_{_{\!}k_{\!}},\xi_{_{\!}k_{\!}-_{\!}1\!},\theta_{_{\!}k\!})\!}(_{\!}Z_{_{\!}k\!})\!\!\! &\propto
&\!\!\!\omega_{_{\!}k_{_{\!}}|_{_{\!}}k_{\!}-_{\!}1{\!}}^{(\!I_{_{\!}k_{\!}},\xi_{_{\!}k_{\!}-_{\!}1\!})}\left[\eta_{_{\!}Z_{_{\!}k}}^{_{\!}(_{\!}\theta_{_{\!}k\!})}(\cdot)\right]^{\!I_{_{\!}k}},  \label{eq:PropBayes_strong1} \\
\eta_{_{\!}Z_{_{\!}k}}^{_{\!}(_{\!}\theta_{_{\!}k\!})}(\ell)\!\!\! &=&\!\!\!\left\langle p_{_{\!}k_{_{\!}}|_{_{\!}}k_{\!}-_{\!}1{\!}}^{_{\!}(_{\!}\xi_{_{\!}k_{\!}-_{\!}1\!})\!}(\cdot ,\ell ),\psi_{_{\!}Z_{_{\!}k}\!}(\cdot ,\ell ;\theta_{_{\!}k\!}
)\right\rangle ,  \label{eq:PropBayes_strong2} \\
p_{k}^{_{\!}(_{\!}\xi_{_{\!}k_{\!}-_{\!}1\!},\theta_{_{\!}k\!})}(x,\ell |Z_k)\!\!\! &=&\!\!\!\frac{p_{_{\!}k_{_{\!}}|_{_{\!}}k_{\!}-_{\!}1{\!}}^{_{\!}(_{\!}\xi_{_{\!}k_{\!}-_{\!}1\!})\!}(x,\ell)\psi_{_{\!}Z_{_{\!}k}\!}(x,\ell;\theta_{_{\!}k\!})}{\eta_{_{\!}Z_{_{\!}k}}^{_{\!}(_{\!}\theta_{_{\!}k\!})}(\ell )}.  \label{eq:PropBayes_strong3}
\end{eqnarray}
\end{proposition}

\begin{figure}[htb]
\centering
\includegraphics[scale=.73]{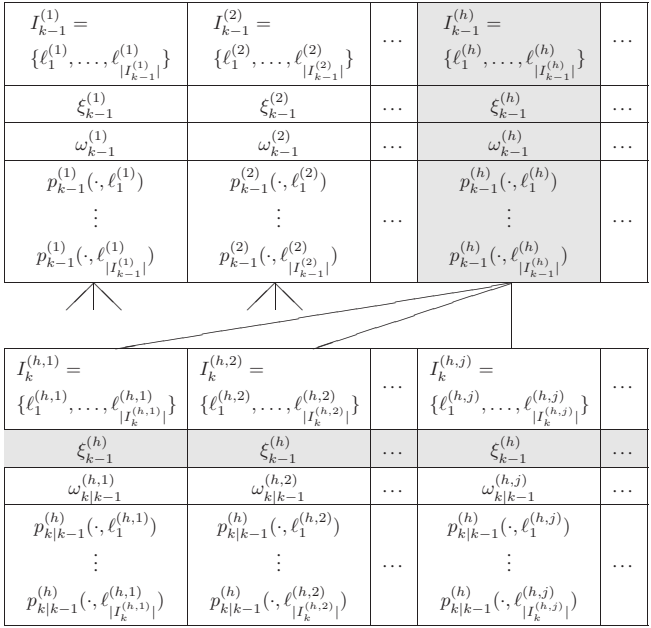}
\caption{The $\protect\delta$-GLMB prediction \cite{VVP_GLMB13}: component $h$ in the prior generates a large set of predicted components with $I_{_{\!}k}^{_{\!}(_{\!}h_{\!},j_{\!})} \subseteq I_{_{\!}k_{\!}-_{\!}1}^{_{\!}(_{\!}h_{\!})\!}\cup\mathbb{B}$, i.e. $j=1$,\ldots, $2^{|_{\!}I_{_{\!}k_{\!}-_{\!}1\!}^{_{\!}(_{\!}h_{\!})\!}|_{\!}+_{\!}|\mathbb{B}|}$, and $\protect\omega_{_{\!}k|k\!-\!1}^{_{\!}(_{\!}h,j_{\!})}\triangleq \protect\omega_{_{\!}k|k\!-\!1}^{{\!}\left({\!}I_{k}^{_{\!}(_{\!}h,j{\!})\!},\protect\xi_{_{\!}k_{\!}-_{\!}1\!}^{_{\!}(_{\!}h_{\!})\!}\right)}$.}
\label{fig:pred_propag}
\end{figure}

\begin{figure}[hbt]
\centering
\includegraphics[scale=.75]{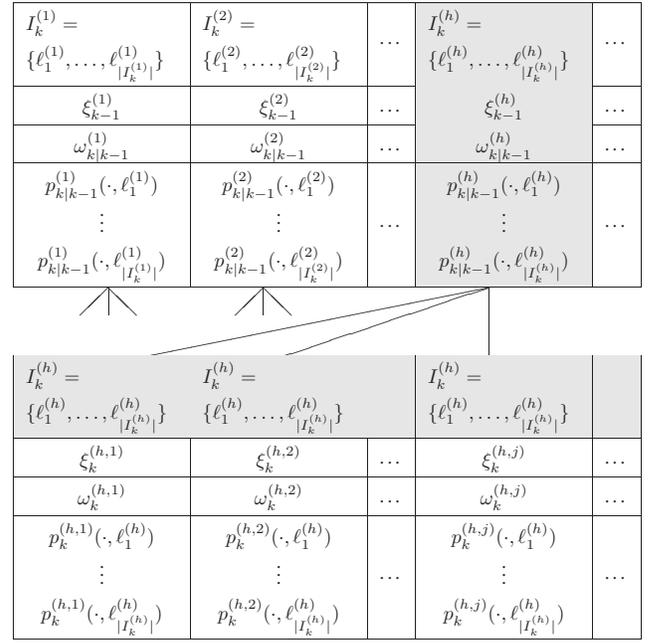}
\caption{The $\protect\delta$-GLMB update \cite{VVP_GLMB13}: component $h$ in the predicted density
generates a (large) set of update components with $\protect\xi_{_{\!}k}^{_{\!}(_{\!}h,j_{\!})}\!\triangleq\!\left(\protect\xi_{_{\!}k_{\!}-_{\!}1_{\!}}^{_{\!}(_{\!}h_{\!})\!},\protect\theta_{_{\!}k\!}^{_{\!}(_{\!}h,j_{\!})\!}\right)$ and weights $\protect\omega_{_{\!}k\!}^{_{\!}(_{\!}h,j_{\!})}\!\triangleq \!\protect\omega_{_{\!}k\!}^{{\!}\left({\!}I_{_{\!}k\!}^{_{\!}(_{\!}h_{\!})\!},\protect\xi_{_{\!}k_{\!}-_{\!}1_{\!}}^{_{\!}(_{\!}h_{\!})\!},\protect\theta_{_{\!}k\!}^{_{\!}(_{\!}h,j_{\!}){\!}}\right)\!}, j=1$,\ldots,$\vert\Theta_{\!I_{_{\!}k}^{_{\!}(_{\!}h_{\!})_{\!}}}\vert$.}
\label{fig:upd_propag}
\end{figure}

The propagations of $\delta$-GLMB components through prediction and update are illustrated in Fig.~\ref{fig:pred_propag} and Fig.~\ref{fig:upd_propag}, respectively. It is clear that the the number of  components grows exponentially with time. Specifically, a component in the filtering density at time $k-1$ generates a large number of predicted components, of which each one in turn produces a new set of multiple $\delta$-GLMB components in the filtering density at time $k$. Hence, it is necessary to reduce the number of $\delta$-GLMB components in both prediction and update densities at every time step. 

The simplest way to truncate a $\delta$-GLMB density is discarding components with smallest weights. The following proposition asserts that this strategy minimizes the $L_{1}$-distance between the true density and the truncated one \cite{VVP_GLMB13}
\begin{proposition}
\label{Prop_L1_error}Let $\left\Vert \mathbf{f}\right\Vert _{1}\triangleq
\int \left\vert \mathbf{f}(\mathbf{X})\right\vert \delta \mathbf{X}$ denote
the $L_{1}$-norm of $\mathbf{f:\mathcal{F}(}\mathbb{X}\mathcal{\times }%
\mathbb{L)}\rightarrow \mathbb{R}$, and for a given $\mathbb{H\subseteq }%
\mathcal{F}(\mathbb{L})\times \Xi $ let 
\begin{equation*}
\mathbf{f}_{\mathbb{H}}\mathbf{(X)}=\Delta (\mathbf{X})\sum\limits_{(I,\xi
)\in \mathbb{H}}\omega ^{(I,\xi )}\delta _{I}(\mathcal{L(}\mathbf{X}))\left[
p^{(\xi )}\right] ^{\mathbf{X}}
\end{equation*}%
be an unnormalized $\delta $-GLMB density. If $\mathbb{T\subseteq H}$ then%
\begin{eqnarray*}
||\mathbf{f}_{\mathbb{H}}-\mathbf{f}_{\mathbb{T}}||_{1}\!\! &\mathbf{=}%
&\!\!\sum\limits_{(I,\xi )\in \mathbb{H-T}}\omega ^{(I,\xi )}, \\
\left\Vert \frac{\mathbf{f}_{\mathbb{H}}}{||\mathbf{f}_{\mathbb{H}}||_{1}}-%
\frac{\mathbf{f}_{\mathbb{T}}}{||\mathbf{f}_{\mathbb{T}}\mathbf{||}_{1}}%
\right\Vert _{1} \!\!&\leq\!\! &2\frac{||\mathbf{f}_{\mathbb{H}}||_{1}-||%
\mathbf{f}_{\mathbb{T}}\mathbf{||}_{1}}{||\mathbf{f}_{\mathbb{H}}||_{1}}.
\end{eqnarray*}
\end{proposition}

\section{Joint prediction and update for the $\delta$-GLMB filter}\label{sec:fast_impl}

In this section, we briefly review the original implementation of the $\delta$-GLMB filter in subsection~\ref{subsec:orig_scheme} and propose a new implementation strategy with joint prediction and update in subsection~\ref{subsec:new_scheme}. 

\subsection{The original implementation}\label{subsec:orig_scheme}

The first implementation of the $\delta$-GLMB filter, detailed in \cite{VVP_GLMB13}, recursively calculates the filtering density by sequentially computing the predicted and update densities at each iteration based on Proposition~\ref{Prop_CK_strong} and Proposition~\ref{PropBayes_strong}. Since direct implementation of equations \eqref{eq:PropCKstrong1} and \eqref{eq:PropBayes_strong0} is difficult due to the sum over supersets in \eqref{eq:PropCKstrongws}, the predicted and update densities are rewritten as \eqref{eq:PropCK_strong3} and \eqref{eq:PropBayes_strong4}, respectively, with 
\[\omega_{_{\!}S}^{_{\!}(_{\!}I_{_{\!}k_{\!}-_{\!}1\!},\xi_{_{\!}k_{\!}-_{\!}1\!})\!}(_{\!}L_{\!})=\left[_{\!}1\!-\!\eta_{_{\!}S}^{_{\!}(_{\!}\xi_{_{\!}k_{\!}-_{\!}1\!})\!}\right]^{\!I_{_{\!}k_{\!}-_{\!}1\!}-_{\!}L\!}\!\left[_{\!}\eta_{_{\!}S}^{_{\!}(_{\!}\xi_{_{\!}k_{\!}-_{\!}1\!})\!}\right]^{\!L\!}.\]
\begin{figure*}[!t]
\normalsize
\setcounter{MYtempeqncnt}{\value{equation}}
\setcounter{equation}{\value{MYtempeqncnt}}
\begin{IEEEeqnarray}{c}
\!\!\!\!\mathbf{\pi}_{_{\!}k_{_{\!}}|_{_{\!}}k_{\!}-_{\!}1\!}(_{\!}\mathbf{X}_{_{\!}}|_{_{\!}}Z_{_{\!}k_{\!}-_{\!}1\!})\!=\!\Delta_{\!}(_{\!}\mathbf{X}_{}\!)\displaystyle\!\!\!\!\!\sum_{I_{_{\!}k_{\!}-_{\!}1}\!,\xi_{_{\!}k_{\!}-_{\!}1}\!, L,J}\!\!\!\!\! 1_{\!\mathcal{F}_{_{\!}}(_{\!}I_{_{\!}k_{\!}-_{\!}1\!})\!}(_{\!}L_{\!})1_{\!\mathcal{F}_{_{\!}}(_{_{\!}}\mathbb{B}_{{\!}})\!}(_{\!}J)\omega_{_{\!}k_{\!}-_{\!}1}^{_{\!}(_{\!}I_{_{\!}k_{\!}-_{\!}1\!},\xi_{_{\!}k_{\!}-_{\!}1\!})}\omega_{_{\!}S}^{_{\!}(_{\!}I_{_{\!}k_{\!}-_{\!}1\!},\xi_{_{\!}k_{\!}-_{\!}1\!})\!}(_{\!}L_{\!})\omega_{_{\!}B\!}(_{\!}J)_{\!}\delta_{_{\!}L\cup J_{_{\!}}}({\!}\mathcal{L}(_{\!}\mathbf{X}_{\!}){\!})\!\!\left[p_{_{\!}k_{_{\!}}|_{_{\!}}k_{\!}-_{\!}1\!}^{_{\!}(\xi_{_{\!}k_{\!}-_{\!}1\!})}\!(_{\!}\cdot_{_{\!}}|_{_{\!}}Z_{_{\!}k_{\!}-_{\!}1\!})_{\!}\right]^{\!\mathbf{X}_{}}\label{eq:PropCK_strong3}\\
\!\!\!\!\!\!\!\!\mathbf{\pi}_{_{\!}k\!}(\mathbf{X}_{\!}\vert_{\!}Z_{_{\!}k\!})\!=\!\frac{\Delta_{\!}(_{\!}\mathbf{X}_{\!})\!\!\!\!\!
\displaystyle\sum_{I_{_{\!}k_{\!}-_{\!}1}\!,\xi_{_{\!}k_{\!}-_{\!}1}\!, L,J,\theta_{_{\!}k\!}} \!\!\!\!\!\! 1_{\!\mathcal{F}_{_{\!}}(_{\!}I_{_{\!}k_{\!}-_{\!}1\!})\!}(_{\!}L^{\!})1_{\!\mathcal{F}_{_{\!}}(_{_{\!}}\mathbb{B}_{{\!}})\!}({\!}J^{_{\!}})1_{_{\!}{\Theta}_{_{\!}L\cup J\!}}(_{\!}\theta_{_{\!}k\!})\omega_{_{\!}k_{\!}-_{\!}1}^{_{\!}(_{\!}I_{_{\!}k_{\!}-_{\!}1\!},\xi_{_{\!}k_{\!}-_{\!}1\!})}\omega_{_{\!}S}^{_{\!}(_{\!}I_{_{\!}k_{\!}-_{\!}1\!},\xi_{_{\!}k_{\!}-_{\!}1\!})\!}(_{\!}L^{\!})\omega_{_{\!}B\!}(_{\!}J^{_{\!}})\!\!\left[_{\!}\eta_{_{\!}Z_{_{\!}k}}^{_{\!}(_{\!}\theta_{_{\!}k\!})\!}\right]^{\!L\cup_{_{\!}}J}\!\!\delta_{_{\!}L\cup J_{_{\!}}}({\!}\mathcal{L}(_{\!}\mathbf{X}_{\!}){\!})\!\!\left[p_{_{\!}k_{_{\!}}}^{_{\!}(_{\!}\xi_{_{\!}k_{\!}-_{\!}1\!},\theta_{_{\!}k\!})}\!(_{\!}\cdot_{_{\!}}|_{_{\!}}Z_{_{\!}k\!})_{\!}\right]^{\!\mathbf{X}_{}}}
{\displaystyle\sum_{I_{_{\!}k_{\!}-_{\!}1}\!,\xi_{_{\!}k_{\!}-_{\!}1}\!, L,J,\theta_{_{\!}k\!}} \!\!\!\!\!\! 1_{\!\mathcal{F}_{_{\!}}(_{\!}I_{_{\!}k_{\!}-_{\!}1\!})\!}(_{\!}L^{\!})1_{\!\mathcal{F}_{_{\!}}(_{_{\!}}\mathbb{B}_{{\!}})\!}({\!}J^{_{\!}})1_{_{\!}{\Theta}_{_{\!}L\cup J\!}}(_{\!}\theta_{_{\!}k\!})\omega_{_{\!}k_{\!}-_{\!}1}^{_{\!}(_{\!}I_{_{\!}k_{\!}-_{\!}1\!},\xi_{_{\!}k_{\!}-_{\!}1\!})}\omega_{_{\!}S}^{_{\!}(_{\!}I_{_{\!}k_{\!}-_{\!}1\!},\xi_{_{\!}k_{\!}-_{\!}1\!})\!}(_{\!}L^{\!})\omega_{_{\!}B\!}(_{\!}J^{_{\!}})\!\!\left[_{\!}\eta_{_{\!}Z_{_{\!}k}}^{_{\!}(_{\!}\theta_{_{\!}k\!})\!}\right]^{\!L\cup_{_{\!}}J}} \label{eq:PropBayes_strong4}
\end{IEEEeqnarray}
\setcounter{equation}{\value{MYtempeqncnt}+2}
\hrulefill
\vspace*{4pt}
\end{figure*}

In the prediction stage \eqref{eq:PropCK_strong3}, each component $(_{\!}I_{_{\!}k_{\!}-_{\!}1\!},\xi_{_{\!}k_{\!}-_{\!}1\!})$ with weight $\omega_{_{\!}k_{\!}-_{\!}1}^{_{\!}(_{\!}I_{_{\!}k_{\!}-_{\!}1\!},\xi_{_{\!}k_{\!}-_{\!}1\!})}$ generates a set of prediction components $(L_{\!}\cup_{\!}J_{\!},\xi_{k_{\!}-_{\!}1})$ with weight
\[\omega_{_{\!}k_{_{\!}}|_{_{\!}}k_{\!}-_{\!}1\!}^{{_{\!}(_{\!}L_{_{\!}}\cup_{_{\!}}J_{\!},\xi_{_{\!}k_{\!}-_{\!}1\!})\!}} = \omega_{_{\!}k_{\!}-_{\!}1}^{_{\!}(_{\!}I_{_{\!}k_{\!}-_{\!}1\!},\xi_{_{\!}k_{\!}-_{\!}1\!})}\omega_{_{\!}S}^{_{\!}(_{\!}I_{_{\!}k_{\!}-_{\!}1\!},\xi_{_{\!}k_{\!}-_{\!}1\!})\!}(_{\!}L^{\!})\omega_{_{\!}B\!}(_{\!}J^{_{\!}}),\]
where $L$ and $J$ represent two disjoint label sets for survival and birth tracks, respectively. Since the weight of the prediction component can be factorized into two factors, $\omega_{_{\!}S}^{_{\!}(_{\!}I_{_{\!}k_{\!}-_{\!}1\!},\xi_{_{\!}k_{\!}-_{\!}1\!})\!}(_{\!}L^{\!})$ and $\omega_{_{\!}B\!}(_{\!}J^{_{\!}})$, which depend on two mutually exclusive sets; truncating the predicted density is performed by solving two separate $K$-shortest path problems for each set of tracks. This is because running only one instance of the $K$-shortest path based on the augmented set of existing and birth tracks generally favours the selection of survival tracks over new births and typically results in poor track initiation.

In the update stage \eqref{eq:PropBayes_strong4}, each prediction component $(L_{\!}\cup_{\!}J_{\!},\xi_{k_{\!}-_{\!}1})$ generates a (large) set of update components $(L_{\!}\cup_{\!}J_{\!},(\xi_{k_{\!}-_{\!}1\!},\theta_{_{\!}k\!}))$. These update components are truncated without having to exhaustively compute all the components by solving a ranked assignment problem.

Although the original two-staged implementation is intuitive and highly
parallelizable, it is has several drawbacks. First, since truncation of the
predicted $\delta $-GLMB density is performed separately from the update based purely on \textit{a priori} knowledge (e.g. survival and birth probabilities), in general, a significant portion of the predicted components would generate
updated components with negligible weights. Hence, computations are wasted
in solving a large number of ranked assignment problems, each of which has
cubic complexity in the number of measurements. Second, it would be very difficult to determine the final approximation error of the truncated filtering density as the implementation involves least three separate truncating processes: one for existing tracks, one for birth tracks, and one for predicted tracks.

In the following subsections, we will introduce the joint prediction and update as a better alternative to the original two-staged approach. The joint strategy eliminates the need for separate prediction truncating procedures, thus involves only one truncation per iteration. Consequently, the new implementation yields considerable computational savings while preserving the filtering performance as well as the parallelizability of the original implementation.

\subsection{The joint prediction and update implementation}\label{subsec:new_scheme}

Instead of computing the filtering density in two steps, the new strategy aims to generate the components of the filtering density in one combined step by formulating a direct relationship between the component of the current filtering density with those of the previous density. Specifically, we will derive a new formulation for $\mathbf{\pi}_{_{\!}k\!}(\mathbf{X}_{\!}\vert_{\!}Z_{_{\!}k\!})$ that does not involve prediction induced variables $L$ and $J$. This can be done via an \textit{extended association map}, denoted by $\tilde{\theta}$, and defined as follows
\begin{definition}
The extended association map is a
function $\tilde{\theta}: \mathbb{L}\cup\mathbb{B} \to \{0,1,\ldots,|Z_{\!k}|, |Z_{\!k}|\!+\!1\}$  such that $\tilde{\theta}{(i)}\!=\!\tilde{\theta}{(i^\prime)}$ for $%
0\!<\!\tilde{\theta}(i)\!<|Z_{\!k}|\!+\!1$ implies $i=i^\prime$.
\end{definition}
\begin{remark}
The new map, in essence, only extends the original map to include a new association, $|Z_{\!k}|\!+\!1$. In particular, $\tilde{\theta}$ is identical to $\theta$ except for non-survival and unconfirmed birth tracks, i.e.
\begin{equation}
\tilde{\theta}(\ell)=
\begin{cases}
\theta(\ell) & \forall\ell \in L \cup J,\\
|Z_{\!k}|+1 & \forall\ell \in (I_{\!k\!-\!1} - L) \cup (\mathbb{B}-J).
\end{cases}
\end{equation}

The image of a set $I \subset \mathbb{L}\cup\mathbb{B}$ through the extended map $\tilde{\theta}_{}$ and the collection of all eligible extended association maps on domain $I$ are denoted by $\tilde{\theta}^{_{\!}}(I)$ and $\tilde{\Theta}_{_{\!}I}$, respectively. 
\end{remark}

Based on the notion of extended association map, the following proposition establishes the direct relationship between two consecutive filtering densities at time $k$ and $k-1$. For simplicity, we assume that target births are modeled by (labeled) multi-Bernoulli RFS's, i.e. $\omega_B(J)=[1-r(\cdot)]^{\mathbb{B}-J}[r(\cdot)]^J$ with $r(\ell)$ denotes the existence probability of track $\ell$.

\begin{proposition}
\label{Prop_joint}
If the multi-target posterior at time $k-1$ is a $\delta $-GLMB of the form \eqref{eq:priorGLMB}
and the set of targets born at the next time is distributed as a labeled multi-Bernoulli RFS,
then the multi-target posterior at the next time is a $\delta $-GLMB given by \allowdisplaybreaks%
\begin{align}\label{eq:PropBayes_strong5}
\!\mathbf{\pi}_{_{\!}k\!}(\mathbf{X}_{\!}\vert_{\!}Z_{_{\!}k\!})= &\, \Delta_{\!}(_{\!}\mathbf{X}_{\!})\!\!\!\!\sum_{I_{_{\!}k_{\!}-_{\!}1}\!,\xi_{_{\!}k_{\!}-_{\!}1}\!, \tilde{\theta}_{_{\!}k}} \!\!\!\! 1_{_{\!}{\tilde{\Theta}}_{_{\!}I_{_{\!}k_{\!}-_{\!}1\!}\cup_{_{\!}} \mathbb{B}\!}}(_{\!}\tilde{\theta}_{_{\!}k\!})\omega_k^{_{\!}(_{\!}I_{_{\!}k_{\!}-_{\!}1\!},\xi_{_{\!}k_{\!}-_{\!}1\!},\tilde{\theta}_{_{\!}k\!})}\times\notag\\
&\qquad\quad1_{_{\!}\{0,\ldots,|_{\!}Z_{\!k\!}|\}\!}(_{\!}\tilde{\theta}_{_{\!}k_{\!}}(_{\!}\mathcal{L}(_{\!}\mathbf{X}_{\!}){\!})\!)\!\!\left[p_{_{\!}k_{_{\!}}}^{_{\!}(_{\!}\xi_{_{\!}k_{\!}-_{\!}1\!},\tilde{\theta}_{_{\!}k\!})}\!(_{\!}\cdot_{_{\!}}|_{_{\!}}Z_{_{\!}k\!})_{\!}\right]^{\!\mathbf{X}_{}}
\end{align}
where $p_{_{\!}k_{_{\!}}}^{_{\!}(_{\!}\xi_{_{\!}k_{\!}-_{\!}1\!},\tilde{\theta}_{_{\!}k\!})}\!(_{\!}\ell_{_{\!}}|_{_{\!}}Z_{_{\!}k\!})_{\!} \equiv p_{_{\!}k_{_{\!}}}^{_{\!}(_{\!}\xi_{_{\!}k_{\!}-_{\!}1\!},{\theta}_{_{\!}k\!})}\!(_{\!}\ell_{_{\!}}|_{_{\!}}Z_{_{\!}k\!})_{\!}, \, \forall\ell\!: \tilde{\theta}_{_{\!}k}(_{\!}\ell^{_{\!}})\!<\!|_{\!}Z_{\!k\!}|\!+\!1$ and
\begin{align}
&\!\omega_k^{_{\!}(_{\!}I_{_{\!}k_{\!}-_{\!}1\!},\xi_{_{\!}k_{\!}-_{\!}1\!},\tilde{\theta}_{_{\!}k\!})}\propto\omega_{_{\!}k_{\!}-_{\!}1}^{_{\!}(_{\!}I_{_{\!}k_{\!}-_{\!}1\!},\xi_{_{\!}k_{\!}-_{\!}1\!})}\!\left[_{\!}\gamma_{_{\!}Z_{_{\!}k}}^{_{\!}(_{\!}\tilde{\theta}_{_{\!}k\!}(\cdot))\!}(\cdot)\right]^{\!I_{_{\!}k_{\!}-_{\!}1\!}\cup_{_{\!}} \mathbb{B}}\\
&\!\gamma_{_{\!}Z_{_{\!}k}}^{_{\!}(_{\!}\tilde{\theta}_{_{\!}k\!}(_{\!}\ell^{_{\!}})_{\!})\!}(_{\!}\ell^{_{\!}})\!=\!\begin{cases}
1_{\!}-_{\!}{\eta}_{_{\!}S}^{_{\!}(_{\!}\xi_{k\!-\!1\!}^{})}\!(_{\!}\ell^{_{\!}}) & \!\forall\ell\!\in\! I_{_{\!}k_{\!}-_{\!}1}\!\!: \tilde{\theta}_{_{\!}k_{\!}}(_{\!}\ell^{_{\!}})\!=\!|Z_{\!k}|\!+\!1, \\
{\eta}_{_{\!}S}^{_{\!}(_{\!}\xi_{k\!-\!1\!}^{})}\!(_{\!}\ell^{_{\!}}){\eta}_{_{\!}Z_{_{\!}k}}^{_{\!}(_{\!}{\theta}_{_{\!}k\!})\!}(_{\!}\ell^{_{\!}}) & \!\forall\ell\!\in\!I_{_{\!}k_{\!}-_{\!}1}\!\!: \tilde{\theta}_{_{\!}k_{\!}}(_{\!}\ell^{_{\!}})\!<\!|Z_{\!k}|\!+\!1, \\ 
1_{\!}-_{\!}r(_{\!}\ell^{_{\!}}) & \!\forall\ell\!\in\!\mathbb{B}\!: \tilde{\theta}_{_{\!}k_{\!}}(_{\!}\ell^{_{\!}})\!=\!|Z_{\!k}|\!+\!1, \\
r(_{\!}\ell^{_{\!}}){\eta}_{_{\!}Z_{_{\!}k}}^{_{\!}(_{\!}{\theta}_{_{\!}k\!})\!}(_{\!}\ell^{_{\!}}) & \!\forall\ell\!\in\!\mathbb{B}\!: \tilde{\theta}_{_{\!}k_{\!}}(_{\!}\ell^{_{\!}})\!<\!|Z_{\!k}|\!+\!1. \label{eq:gamma_def}
\end{cases}
\end{align}
\end{proposition}

We now proceed to detail an efficient implementation of the $\delta$-GLMB filter based on the result in Proposition~\ref{Prop_joint}. Let the sets of existing tracks, birth tracks, and measurements be enumerated by $I^{(h)}_{k\!-\!1}\!=\!\{\ell_{1}\!,\ldots\!, \ell_{N}\}$, $\mathbb{B}\!=\!\{\ell_{N\!+\!1}\!,\ldots\!, \ell_{P}\}$, and $Z_{\!k}\!=\!\{z_1, \ldots , z_M\}$, respectively. Given the $h$-th component $(I^{(h)}_{k\!-\!1}, \xi^{(h)}_{k\!-\!1})$, the objective is to find $T^{(h)}$ extended associations $\left\{\tilde{\theta}_{_{\!}k}^{_{\!}(_{\!}h_{\!},j_{\!})}\right\}_{_{\!}j=1}^{_{\!}T^{(h)}} \subset \tilde{\Theta}_{I^{(h)}_{k\!-\!1}\cup\mathbb{B}}$ that produce highest update weights.

Denote by $S_{_{\!}j}$ a $P \times (M+2P)$ matrix whose entries are either $1$ or $0$ such that the sum of each row is exactly $1$ while the sum of each column is at most $1$. The matrix $S_{_{\!}j}$ is called an \textit{assignment matrix} since it represents a valid extended association map $\tilde{\theta}_{_{\!}k}^{_{\!}(_{\!}h_{\!},j_{\!})} \in \tilde{\Theta}_{I^{(h)}_{k\!-\!1}\cup\mathbb{B}}$. Hence, finding the desirable extended associations is equivalently translated to finding the corresponding assignment matrices. The simplest way to determine these matrices without exhaustingly computing all of the update weights is to assign an appropriate cost for each assignment matrix and then rank these matrices in non-decreasing order of their costs via Murty's algorithm.

\begin{figure*}[!t]
\centering
\includegraphics[scale=.75,angle=270]{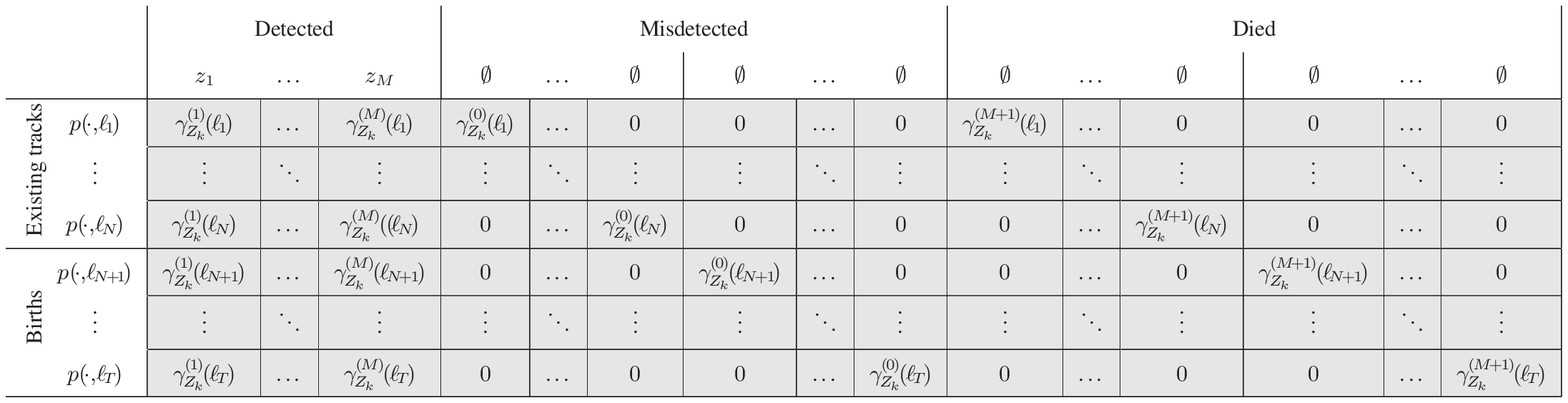}
\caption{The matrix $\Gamma^{_{\!}(\!h\!)}_{{\!}Z_{_{\!}k}}$ whose entry $\gamma_{_{\!}Z_{_{\!}k}\!}^{_{\!}(_{\!}m_{\!})\!}(_{\!}\ell_{\!n}\!)$ represents the likelihood that $\tilde{\theta}_{_{\!}k}^{_{\!}(_{\!}h_{\!},j_{\!})\!}(_{\!}\ell_{_{\!}n\!})=m$ for any valid extended association $\tilde{\theta}_{_{\!}k}^{_{\!}(_{\!}h_{\!},j_{\!})\!} \in \tilde{\Theta}_{_{\!}I^{_{\!}(_{\!}h_{\!})}_{_{\!}k_{\!}-_{\!}1}\cup\mathbb{B}}$. The cost matrix $C^{_{\!}(\!h\!)}_{_{\!}Z_{_{\!}k}}$ is formed by taking negative logarithm on $\Gamma^{_{\!}(\!h\!)}_{_{\!}Z_{_{\!}k}}$, i.e. $C^{_{\!}(\!h\!)}_{_{\!}Z_{_{\!}k}}=-\log\left(\Gamma^{_{\!}(\!h\!)}_{_{\!}Z_{_{\!}k}}\right)$.}
\label{fig:contr_mat}
\end{figure*}

Let $\Gamma^{_{\!}(\!h\!)}_{{\!}Z_{_{\!}k}}$ be a matrix whose $(n,m)$ entry, with $1\leq n \leq P$ and $1\leq m \leq M+2P$, is defined as follows. 
\begin{equation}\label{eq:Gamma_big}
\Gamma^{_{\!}(\!h\!)}_{{\!}Z_{_{\!}k}}(n,m)=\begin{cases}
\gamma_{Z_k}^{(m)}(\ell_n) & m\leq M,\\
\gamma_{Z_k}^{(0)}(\ell_n) & m=M+n,\\
\gamma_{Z_k}^{(M+1)}(\ell_n) & m=M+P+n,\\
0 & \text{otherwise}.
\end{cases}
\end{equation}
As depicted in Fig.~\ref{fig:contr_mat}, the matrix $\Gamma^{_{\!}(\!h\!)}_{{\!}Z_{_{\!}k}}$ contains all possible values of $\gamma_{_{\!}Z_{_{\!}k}}^{_{\!}(_{\!}\tilde{\theta}_{_{\!}k}^{_{\!}(_{\!}h_{\!},j_{\!})\!}(_{\!}\ell^{_{\!}})_{\!})\!}(\ell)$ that any valid extended associations $\tilde{\theta}_{_{\!}k}^{_{\!}(_{\!}h_{\!},j_{\!})\!}$ can generate. 
Intuitively speaking, each entry of $\Gamma^{_{\!}(\!h\!)}_{{\!}Z_{_{\!}k}}$ is an indicator of how likely an extended association is assigned to a track; hence, the matrix $C_{_{\!}Z_{\!k}}^{_{_{\!}}(_{\!}h_{\!})}=-\log\left(\Gamma^{_{\!}(\!h\!)}_{{\!}Z_{_{\!}k}}\right)$ is called the cost matrix and  cost function for a particular assignment matrix $S_{_{\!}j}$ is given by
\begin{equation}\label{eq:cost_func}
f_c(S_{_{\!}j}) = -\text{tr}\left(S_{_{\!}j}^{T}C_{_{\!}Z_{\!k}}^{_{_{\!}}(_{\!}h_{\!})}\right),
\end{equation}
It is straightforward to show that the (unnormalized) weight of the $\tilde{\theta}_{_{\!}k}^{_{\!}(_{\!}h_{\!},j_{\!})}$-generated component in the filtering density at time $k$ is
\begin{equation}\label{eq:joint_wk_1}
\omega_k^{_{\!}(_{\!}I_{_{\!}k_{\!}-_{\!}1\!},\xi_{_{\!}k_{\!}-_{\!}1\!},\tilde{\theta}_{_{\!}k}^{_{\!}(_{\!}h_{\!},j_{\!})\!})}\!\propto \exp\left[-f_c(S_{_{\!}j})\right].
\end{equation}

Based on \eqref{eq:cost_func} and \eqref{eq:joint_wk_1}, an implementation of the $\delta$-GLMB filter with joint prediction and update is given in Algorithm~\ref{alg:joint_Murty}, where the subroutine $\mathtt{ranked\_assignment}$ uses Murty's algorithm \cite{Murty68} to generate a sequence of $T^{(h)}$ assignment matrices that yield lowest costs (or equivalently, highest update weights) without exhaustively navigating the whole assignment space.
\renewcommand{\algorithmicrequire}{\textbf{Inputs:}} \renewcommand{%
\algorithmicensure}{\textbf{Outputs:}}
\begin{algorithm}[htb]
\caption{$\delta$-GLMB joint prediction and update}
\label{alg:joint_Murty}
\begin{algorithmic}[1]
    \Require $\left\{I_{_{\!}k\!-\!1\!}^{_{\!}(_{\!}h_{\!})},\xi_{_{\!}k\!-\!1\!}^{_{\!}(_{\!}h_{\!})},\omega_{_{\!}k\!-\!1\!}^{_{\!}(_{\!}h_{\!})},p_{_{\!}k\!-\!1\!}^{_{\!}(_{\!}h_{\!})},T^{_{_{\!}}(_{\!}h_{\!})}\right\}_{\!h=1\!}^{\!H}, Z_{_{\!}k}$ 
    \Ensure $\left\{I_{_{\!}k}^{_{\!}(_{\!}h_{\!},j_{\!})\!},\xi_{_{\!}k}^{_{\!}(_{\!}h_{\!},j_{\!})\!},\omega_{_{\!}k}^{_{\!}(_{\!}h_{\!},j_{\!})\!},p_{_{\!}k}^{_{\!}(_{\!}h_{\!},j_{\!})\!}\right\}_{_{\!}(_{\!}h_{\!},j_{\!})=(_{\!}1_{\!},1_{\!})}^{_{\!}(_{\!}H_{\!},T^{_{_{\!}}(_{\!}h_{\!})\!})}$
    \For{$h \gets 1, H$}
    \State compute $\Gamma^{_{\!}(_{\!}h_{\!})}_{{\!}Z_{_{\!}k}}$ according to \eqref{eq:gamma}
    \State $C_{{\!}Z_{_{\!}k}}^{_{\!}(_{\!}h_{\!})} \gets -\log\!\left(\Gamma^{_{\!}(\!h\!)}_{{\!}Z_{_{\!}k}}\right)$
    \State $\left(\!I_{_{\!}k}^{_{\!}(_{\!}h_{\!},j_{\!})\!},\xi_{_{\!}k}^{_{\!}(_{\!}h_{\!},j_{\!})\!}\!\right)\gets\mathsf{ranked\_assignment}(C_{{\!}Z_{_{\!}k}}^{_{\!}(_{\!}h_{\!})\!},T^{_{_{\!}}(_{\!}h_{\!})\!})$
    \For{$j \gets 1, T^{_{_{\!}}(_{\!}h_{\!})\!}$}
    \State compute $\omega_{_{\!}k}^{_{\!}(_{\!}h_{\!},j_{\!})\!}$ according to \eqref{eq:joint_wk_1}
    \State compute $p_{_{\!}k}^{_{\!}(_{\!}h_{\!},j_{\!})\!}$ according to \eqref{eq:PropBayes_strong3}
    \EndFor 
    \EndFor
    \State normalize weights $\left\{\omega_{_{\!}k}^{_{\!}(_{\!}h_{\!},j_{\!})\!}\right\}_{_{\!}(_{\!}h_{\!},j_{\!})=(_{\!}1_{\!},1_{\!})}^{_{\!}(_{\!}H_{\!},T^{_{_{\!}}(_{\!}h_{\!})\!})}$
\end{algorithmic}
\end{algorithm}

Similar to the original implementation, the joint prediction and update also operates independently on each components in the filtering density, thereby is highly parallelizable.

\subsection{Stochastic simulation approach to extended data associations}
\label{subsec:fast_assginment} 
In multi-target tracking, truncating procedures based on the original 
Murty's algorithm with complexity $O\!\left( \!T\left\vert
Z\right\vert ^{4}\!\right) $, where $T$ is the number of assignments and $%
|Z| $ is the number of the measurements, have been proposed in \cite%
{DanchickNewnam93}, \cite{CoxMiller95}, \cite{CoxHingorani96}. More
efficient algorithms with $O\!\left( \!T\left\vert Z\right\vert
^{3}\!\right) $ complexity have been proposed in \cite{Milleretal97}, \cite%
{Pascoaletal03}, \cite{Pedersenetal08}, with the latter showing better
efficiency for large $\left\vert Z\right\vert $. The main drawback of these
approaches is that a significant amount of computation is used to sort the
data associations in a particular order despite the fact that order is
effectively discarded after the update. Furthermore, the number of desired components must be predetermined, generally by a large enough number to
capture all important associations. If the number of significant components in
the filtering density is much smaller than the chosen threshold, many insignificant
components are generated that waste a lot of computation at the next
iteration. Conversely if the number of significant component exceeds the
chosen threshold, the filtering performance will likely degrades in subsequent iterations.
Nonetheless a deterministic polynomial time solution is thus appealing in
the sense that convergence and reproducibility is guaranteed without having
to enumerate all possible solutions.

In this subsection, we propose an alternative to the ranked assignment
based solution. Our proposed solution is based on stochastic simulation or
Markov Chain Monte Carlo methods via a Gibbs sampler which directly address
the above mentioned drawbacks. Conceptually, instead of ranking extended
associations in non-increasing order of their weights, each extended
association is treated as a realization of a (discrete) random variable,
where the probability of each extended association is proportional to the weight of its associated $\delta$-GLMB component 
in the next filtering density. Candidate extended associations are then generated by
sampling from this discrete distribution. Extended associations with high
weights are chosen more often than those with low weights in a
statistically consistent manner. Consequently these samples of extended
associations are more statistically diverse than those from obtained from a
deterministic approach such as the ranked optimal assignment.

Using the same enumeration for tracks and measurement as in the previous section, a valid extended association map $\tilde{\theta}$ for each component $(I_{k\!-\!1}^{(h)},\xi _{k\!-\!1}^{(h)})$ can be represented as a vector ${\tilde{\theta}}=\left[ \tilde{\theta}(\ell _{1}),...,\tilde{\theta}(\ell_{P})\right]^T \in \{0,1,\ldots,M\!+\!1\}^{P}$. The key idea of the stochastic based approach is that $\tilde{\theta}$ can be considered as realizations of a random variable in the space $\tilde{\Theta}_{_{\!}I_{k\!-\!1}^{(h)}\cup\mathbb{B}}$ with the following distribution
\begin{equation}\label{eq:thetajoint_dis}
\mathbf{\pi}(\tilde{\theta})\propto {1}_{_{\!}\tilde{\Theta}_{_{\!}I_{k\!-\!1}^{(h)}\cup\mathbb{B}}}(\tilde{\theta})\omega _{k}^{(I_{k-1}^{(h)},\xi_{k-1}^{(h)},\tilde{\theta})}
\end{equation}%
where
\allowdisplaybreaks 
\begin{align}
{1}_{_{\!}\tilde{\Theta}_{_{\!}I_{k\!-\!1}^{(h)}\cup\mathbb{B}}}(\tilde{\theta})&= \prod_{i=1}^P \left[1-1_{\mathcal{M}\setminus\{\tilde{\theta}_{_{\!}}(_{\!}\ell_{\!i\!})\}\!}(_{\!}\tilde{\theta}_{_{\!}}(_{\!}\ell_{\!i\!})_{\!})\right],\label{eq:theta_req}\\
\omega _{k}^{(I_{k-1}^{(h)},\xi_{k-1}^{(h)},\tilde{\theta})}&\propto \omega
_{k\!-\!1}^{(I_{k\!-\!1}^{(h)},\xi _{k\!-\!1}^{(h)})}\!\left[\gamma_{_{\!}Z_{_{\!}k}\!}^{_{\!}(_{\!}\tilde{\theta}(\cdot)_{\!})\!}(\cdot )\right] ^{\!I_{k\!-\!1}\cup \mathbb{B}},\label{eq:omeg_theta}
\end{align}%
with $\mathcal{M}=\{1,\ldots,M\}\cap\{\tilde{\theta}_{_{\!}}(_{\!}\ell_{\!1\!}),\ldots,\tilde{\theta}_{_{\!}}(_{\!}\ell_{\!P\!})\}$ and $\gamma_{_{\!}Z_{_{\!}k}\!}^{_{\!}(_{\!}\tilde{\theta}(\ell)_{\!})\!}(\ell)$ is given in \eqref{eq:gamma_def}. Thus
the probability of a valid extended association is proportional to the
weight of the corresponding $\delta$-GLMB component in the next filtering density while zero
probability is allocated to extended associations which do not
satisfy the constraint that each measurement is assigned to at most
one track.

However, sampling directly from the distribution \eqref{eq:thetajoint_dis} is very difficult since we cannot exhaustively compute all of the values of $\omega _{k}^{(I_{k-1}^{(h)},\xi_{k-1}^{(h)},\tilde{\theta})}$. A common solution to this kind of problem is to use Markov Chain Monte Carlo (MCMC) methods such as the Gibbs sampler to obtain samples from \eqref{eq:thetajoint_dis} without having to directly compute $\omega _{k}^{(I_{k-1}^{(h)},\xi_{k-1}^{(h)},\tilde{\theta})}$. The Gibbs sampler is a very efficient method to sample a difficult distribution if its conditional marginals can be computed in a simple closed form \cite{Geman_Gibbs84,Cassella_Gibbs92} with proven convergence under generally standard assumptions \cite{Frigessi_Gibbs93,Roberts_Gibbs94}.

The main theoretical contribution in this section is stated in the following proposition, which allows conditional marginals to be computed via the entries of the matrix $\Gamma^{_{\!}(\!h\!)}_{{\!}Z_{_{\!}k}}$.
\begin{proposition}\label{marg_cond}
Denote by $\tilde{\theta}_{_{\!}n}$ the $n$-th element of $\tilde{\theta}$, i.e. $\tilde{\theta}_{_{\!}n}=\tilde{\theta}_{_{\!}}(_{\!}\ell_{_{\!}n_{\!}})$, and $\tilde{\theta}_{\bar{n}\!}$ all the other elements except $\tilde{\theta}_{_{\!}n}$, i.e. $\tilde{\theta}_{\bar{n}\!}=[\tilde{\theta}_{_{\!}1:n_{\!}-_{\!}1\!},\tilde{\theta}_{_{\!}n_{\!}+_{\!}1:P\!}]^T$. Then, the conditional marginal $\mathbf{\pi}^{_{\!}}(_{\!}\tilde{\theta}_{_{\!}n_{\!}}|_{_{\!}}\tilde{\theta}_{\bar{n}\!})$ is given by
\begin{equation}\label{eq:marg_cond}
\mathbf{\pi}^{_{\!}}(_{\!}\tilde{\theta}_{_{\!}n_{\!}}|_{_{\!}}\tilde{\theta}_{\bar{n}\!}) \propto \prod_{i=1}^P\left[_{\!}1_{\!}-_{\!}1_{_{\!}\mathcal{M}\setminus\{\tilde{\theta}_{_{\!}i}\}\!}(_{\!}\tilde{\theta}_{_{\!}i\!})\right]\!\gamma_{_{\!}Z_{_{\!}k}}^{_{\!}(_{\!}\tilde{\theta}_{_{\!}n\!})\!}(_{\!}\ell_{_{\!}n\!}).
\end{equation}
\end{proposition}
\begin{remark}
Propostion~\ref{marg_cond} simply states that the conditional probability $\mathbf{\pi}^{_{\!}}(_{\!}\tilde{\theta}_{_{\!}n_{\!}}|_{_{\!}}\tilde{\theta}_{\bar{n}\!})$ are zero when $\tilde{\theta}_{_{\!}n_{\!}}$ is inconsistent with $\tilde{\theta}_{_{\!}m_{\!}}$, thus ensuring that each measurement can be assigned to at most on track. The conditional probabilities are otherwise proportional to $\gamma_{_{\!}Z_{_{\!}k}}^{_{\!}(_{\!}\tilde{\theta}_{_{\!}n\!})\!}(_{\!}\ell_{_{\!}n\!})$.
\end{remark}

With these conditionals the pseudo code for $\delta $-GLMB filtering with
the Gibbs sampler is given in Algorithm~\ref{alg:Gibbs} where the Gibbs
sampler is used directly in place of the ranked optimal assignment. 
\begin{algorithm}[htb]
\caption{Gibbs sampling of $\delta$-GLMB assignments}
\label{alg:Gibbs}
\begin{algorithmic}[1]
    \Require $I_{_{\!}k_{\!}-_{\!}1\!}^{_{\!}(_{\!}h_{\!})\!}\cup\mathbb{B},\xi_{_{\!}k_{\!}-_{\!}1\!}^{_{\!}(_{\!}h_{\!})},p_{_{\!}k_{\!}-_{\!}1\!}^{_{\!}(_{\!}h_{\!})},T^{_{\!}(_{\!}h_{\!})},B^{_{\!}(_{\!}h_{\!})},Z_{_{\!}k}$
    \Ensure $\left\{I_{_{\!}k}^{_{\!}(_{\!}h_{\!},{j}_{\!})\!},\xi_{_{\!}k}^{_{\!}(_{\!}h_{\!},{j}_{\!})\!}\right\}_{\!{j}_{\!}=_{\!}1}^{_{\!}T^{_{\!}(_{\!}h_{\!})}}$
    \State $\left\{I_{_{\!}k}^{_{\!}(_{\!}h_{\!},\tilde{j})\!},\xi_{_{\!}k}^{_{\!}(_{\!}h_{\!},\tilde{j})\!}\right\} \gets \emptyset$
    \State compute $\Gamma^{_{\!}(_{\!}h_{\!})}_{{\!}Z_{_{\!}k}}$ according to \eqref{eq:Gamma_big}
    \For{$j \gets 1, T^{_{\!}(_{\!}h_{\!})}$}
    \State Initialize $[\tilde{s}_{_{\!}1\!},\ldots,\tilde{s}_{_{\!}P}]^T\gets\left[\tilde{\theta}_{_{\!}k}^{_{\!}(_{\!}h,0_{\!})\!}(_{\!}\ell_{_{\!}1\!}), \ldots,%
\tilde{\theta}_{_{\!}k}^{_{\!}(_{\!}h,0_{\!})\!}(_{\!}\ell_{_{\!}P\!})\right]^T$
    \For{$t \gets 1, B^{_{\!}(_{\!}h_{\!})}$}
    \For{$i \gets 1, P$}
    \State compute $\mathbf{\pi}^{_{\!}}(_{\!}s_{_{\!}i_{\!}}|_{_{\!}}\tilde{s}_{_{\!}1:i_{\!}-_{\!}1\!},\tilde{s}_{_{\!}i_{\!}+_{\!}1:P\!})$ according to  \eqref{eq:marg_cond}
    \State $\tilde{\theta}_{_{\!}k}^{_{\!}(_{\!}h_{\!},t_{\!})\!}(_{\!}\ell_{_{\!}i\!}) \sim \mathbf{\pi}^{_{\!}}(_{\!}s_{_{\!}i_{\!}}|_{_{\!}}\tilde{s}_{_{\!}1:i_{\!}-_{\!}1\!},\tilde{s}_{_{\!}i_{\!}+_{\!}1:P\!})$
    \State $\tilde{s}_{_{\!}i} \gets \tilde{\theta}_{_{\!}k}^{_{\!}(_{\!}h_{\!},t_{\!})\!}(_{\!}\ell_{_{\!}i\!})$
    \EndFor
    \EndFor
    \State $I_{_{\!}k}^{_{\!}(_{\!}h_{\!},j_{\!})} \gets \left\{\ell\in I_{_{\!}k_{\!}-_{\!}1\!}^{_{\!}(_{\!}h_{\!})\!}\cup\mathbb{B}: \tilde{\theta}_{_{\!}k}^{_{\!}(_{\!}h,B^{_{\!}(_{\!}h_{\!})\!})\!}(_{\!}\ell)<M_{\!}+_{\!}1\right\}$
    \State $\xi_{_{\!}k}^{_{\!}(_{\!}h_{\!},{j}_{\!})\!}\gets\left[\xi_{_{\!}k_{\!}-_{\!}1\!}^{_{\!}(_{\!}h_{\!})},\tilde{\theta}_{_{\!}k}^{_{\!}(_{\!}h,B^{_{\!}(_{\!}h_{\!})\!})\!}\right]$
    \State $\left\{I_{_{\!}k}^{_{\!}(_{\!}h_{\!},\tilde{j})\!},\xi_{_{\!}k}^{_{\!}(_{\!}h_{\!},\tilde{j})\!}\right\}_{\!\tilde{j}}\gets\left\{I_{_{\!}k}^{_{\!}(_{\!}h_{\!},\tilde{j})\!},\xi_{_{\!}k}^{_{\!}(_{\!}h_{\!},\tilde{j})\!}\right\}_{\!\tilde{j}_{\!}}\cup \left(I_{_{\!}k}^{_{\!}(_{\!}h_{\!},{j}_{\!})\!},\xi_{_{\!}k}^{_{\!}(_{\!}h_{\!},{j}_{\!})\!}\right)$
    \EndFor
\end{algorithmic}
\end{algorithm}
In order to produce one sample, Algorithm~\ref{alg:Gibbs} requires a Gibbs
sequence of length $B^{_{\!}(_{\!}h_{\!})}$, starting from an arbitrary
initialization $\left[\tilde{\theta}_{_{\!}k}^{_{\!}(_{\!}h,0_{\!})\!}(_{\!}\ell_{_{\!}1\!}), \ldots,%
\tilde{\theta}_{_{\!}k}^{_{\!}(_{\!}h,0_{\!})\!}(_{\!}\ell_{_{\!}P\!})\right]^T$, to be generated.
Alternatively, we can sample a long Gibbs sequence of length $%
B^{_{\!}(_{\!}h_{\!})}\times T^{_{\!}(_{\!}h_{\!})}+1$ and then extract every $B^{_{\!}(_{\!}h_{\!})}$%
-th sample \cite{Geyer92}. The length of the Gibbs
sequence, roughly speaking, depends on the convergence rate of the Gibbs sampler and the
distance from the initial point to the true sample space. If we start with a
good initialization right in the true sample space, we can use all the
samples from the Gibbs sequence \cite{George_Gibbs92}. In practice, one
example of good initialization that allows us to use all of the samples from
the resulting Gibbs sequence is the optimal assignment, which can be
obtained via either Munkres \cite{Munkres57} or Jonker-Volgenant algorithm 
\cite{Jonker87}. Otherwise, we can start with all zeros assignment (i.e. all
tracks are misdetected) that is also valid sample and requires no additional
computation.

In terms of computational complexity, sampling from a discrete distribution
is linear with the weight's length \cite{Devroye86}, therefore the total
complexity of the Gibbs sampling procedure presented in Algorithm~\ref%
{alg:Gibbs} is $O(T^{_{\!}(_{\!}h_{\!})}B^{_{\!}(_{\!}h_{\!})}P(M+2P))$. In comparison, the
fastest ranked optimal assignment algorithm is $O(T^{_{\!}(_{\!}h_{\!})}(M+2P)^{3})$ 
\cite{Pascoaletal03,Pedersenetal08}. For general multi-target tracking
problems in practice, we usually have $B^{_{\!}(_{\!}h_{\!})}, P \ll (M+2P)$, thus the Gibbs sampling algorithm will generally be much faster than the
ranked assignment given the same $T^{_{\!}(_{\!}h_{\!})}$.

\section{Simulation}

\label{sec:sim} In this section we first compare the performance of the
joint prediction and update approach with its traditional separated
counterpart, both employ the ranked assignment algorithm for fair
comparison. Then, we illustrate the superior performance of the Gibbs
sampler based truncation to the conventional ranked assignment via a
difficult tracking scenario with low detection probability and very high
clutter rate.

\begin{figure}[!t]
\centering
\includegraphics[scale=.55]{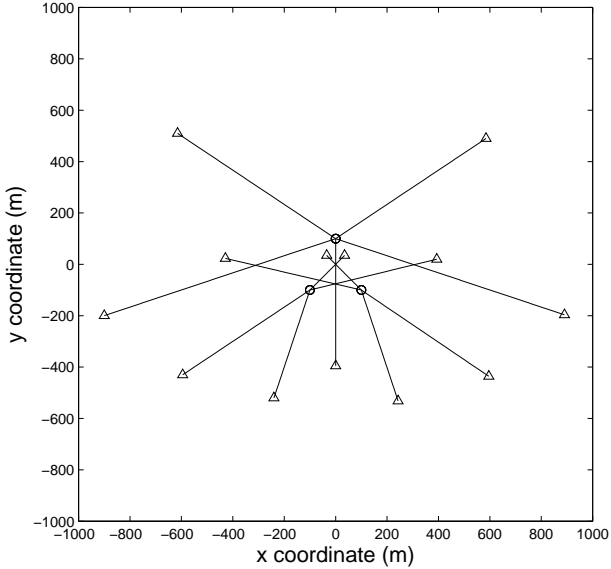}
\caption{Multiple trajectories in the $xy$ plane. Start/Stop positions for
each track are shown with $\circ$/$\bigtriangleup$.}
\label{fig:traj}
\end{figure}

The first numerical example is based on a scenario adapted from \cite%
{VVP_GLMB13} in which a varying number targets travel in straight paths and
with different but constant velocities on the two dimensional region $[-1000,1000]m \times [-1000,1000]m $. The duration of the scenario is $K = 100s$. There is a
crossing of 3 targets at the origin at time $k = 20$, and a crossing of two
pairs of targets at position $(\pm 300,0)$ at time $k = 40$. The region and
tracks are shown in Figure~\ref{fig:traj}. 

The kinematic target state is a vector of planar position and velocity $%
x_k=[p_{x\!,k},p_{y\!,k},\dot{p}_{x\!,k},\dot{p}_{y\!,k}]^T$ . Measurements
are noisy vectors of planar position only $z_k = [z_{x\!,k} ,z_{y\!,k} ]^T$
. The single-target state space model is linear Gaussian according to
transition density $f_{k|k-1}\!(x_k |x_{k-1} ) = \mathcal{N}(x_k;F_k
x_{k-1},Q_k)$ and likelihood $g_k\!(z_k |x_k ) = \mathcal{N}(z_k;H_k
x_k,R_k) $ with parameters 
\begin{align*}
F_k=%
\begin{bmatrix}
I_2 & \Delta I_2 \\ 
0_2 & I_2%
\end{bmatrix}
& \quad Q_k=\sigma_\nu^2%
\begin{bmatrix}
\frac{\Delta^4}{4}I_2 & \frac{\Delta^3}{2}I_2 \\ 
\frac{\Delta^2}{3}I_2 & \Delta^2I_2%
\end{bmatrix}
\\
H_k =%
\begin{bmatrix}
I_2 & 0_2%
\end{bmatrix}
& \qquad R_k=\sigma_\epsilon^2I_2
\end{align*}
where $I_n$ and $0_n$ denote the $n \times n$ identity and zero matrices
respectively, $\Delta = 1s$ is the sampling period, $\sigma_\nu = 5m/s^2$
and $\sigma_\epsilon = 10m$ are the standard deviations of the process noise
and measurement noise. The survival probability is $p_S,k$ = 0.99 and the
birth model is a Labeled Multi-Bernoulli RFS with parameters $\pi_B ={%
r^{\!(\!i\!)}_B\!,p^{\!(\!i\!)}_B}^3_{\!i=1}$ where $r^{\!(\!i\!)}_B= 0.04$
and $p^{\!(\!i\!)}_B\!(x) = \mathcal{N}(x;m^{\!(\!i\!)}_B,P_B )$ with $%
m^{\!(\!1\!)}_B= [ 0,0,100,0 ]^T, m^{\!(\!2\!)}_B= [ -100,0,-100,0]^T, m^{\!(\!3\!)}_B= [ 100,0,-100,0]^T, P_B = \text{diag}([ 10, 10, 10, 10 ]^T)^2$ . The detection
probability is $p_{D,k} = 0.88$ and clutter follows a Poisson RFS with an
average intensity of $\lambda_c = 6.6 \times 10^{\!-\!5}m^{\!-\!2}$
giving an average of $66$ false alarms per scan.

\begin{figure}[htb]
\centering
\includegraphics[scale=.51]{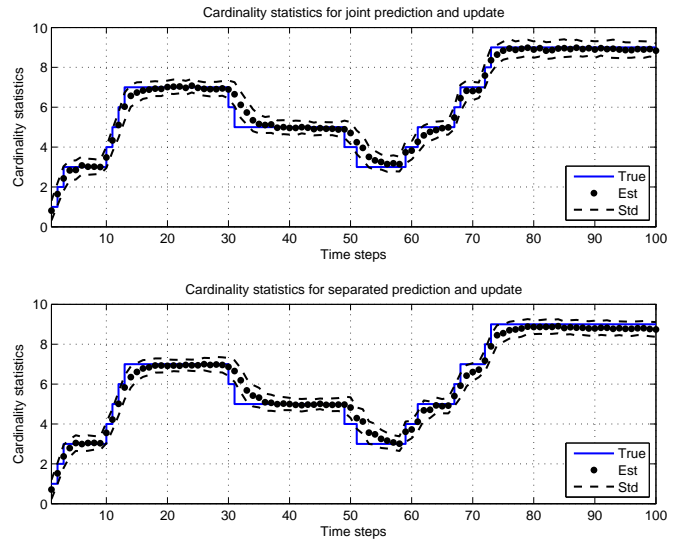}
\caption{Cardinality statistics with traditional ranked assignment
truncation.}
\label{fig:card}
\end{figure}

\begin{figure}[htb]
\centering
\includegraphics[scale=.51]{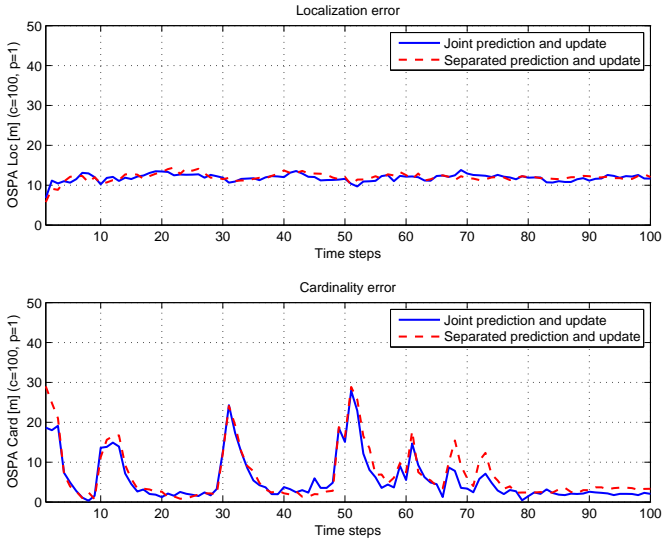}
\caption{OSPA components with traditional ranked assignment truncation.}
\label{fig:ospa_sep}
\end{figure}

First, we compare the performance of the traditional separated and the
proposed joint prediction and update approaches. For a fair comparison, both
approaches are capped to the same maximum components. Results are shown over
100 Monte Carlo trials. Figures~\ref{fig:card} shows the mean and standard
deviation of the estimated cardinality versus time. Figures~\ref%
{fig:ospa_comb} and \ref{fig:ospa_sep} show the OSPA distance \cite{SVV08}
and its localization and cardinality components for $c = 100m$ and $p = 1$. 

\begin{figure}[htb]
\centering
\includegraphics[scale=.51]{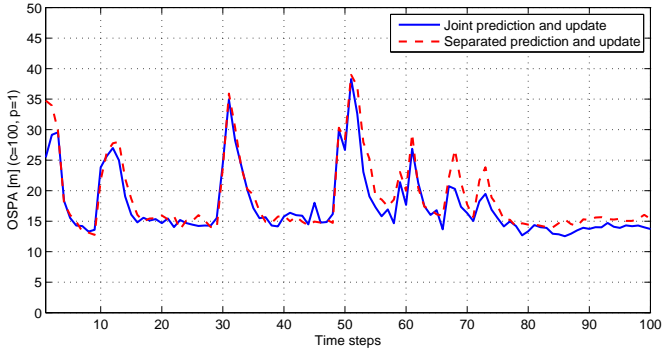}
\caption{OSPA distance with traditional ranked assignment truncation.}
\label{fig:ospa_comb}
\end{figure}

It can be seen that both approaches estimate the cardinality equally well.
Similarly, in terms of OSPA distance, the performance of the two approach is
virtually the same.

\begin{figure}[!t]
\centering
\includegraphics[scale=.5]{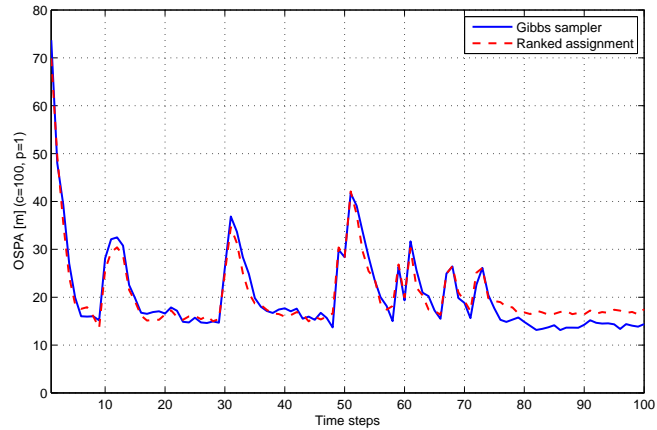}
\caption{OSPA distance comparison between the Gibbs sampler implementation
versus the ranked assignment implementation ($c =100m, p=1$).}
\label{fig:OSPA_100}
\end{figure}

Second, we demonstrate the fast implementation via the Gibbs sampler. In
this example, we keep all parameters the same as in the previous example
except that the clutter rate is now increased to average $100$ false alarms
per scan. The performance of the Gibbs sampler implementation is compared
with that of a ranked assignment based implementation with the same maximum
number of posterior hypotheses. The average OSPA distances over 100 Monte
Carlo trials are presented in Fig.~\ref{fig:OSPA_100}. 

\begin{figure}[t]
\centering
\includegraphics[scale=.5]{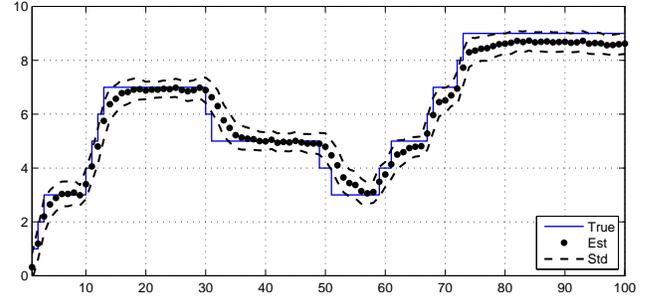}
\caption{Cardinality statistics for ranked assignment implementation.}
\label{fig:card_Murty}
\end{figure}

It is obvious that the Gibbs sampler has a better OSPA from around time $%
k=75 $ onward. The reason is in difficult scenario (e.g. high clutter rate,
low detection probability), if the number of existing targets are high the
Gibbs sampling technique is expected to pick up the new born target better
than the ranked assignment algorithm given the same number of
samples/hypotheses due to its randomized behaviour. This is clearly
illustrated in the cardinality statistics for both approaches in Fig.~\ref%
{fig:card_Gibbs} and Fig.~\ref{fig:card_Murty}. As expected, however, the
joint approach averaged run time is significantly lower that that of the
original approach. Reductions in execution time of 1 to 2 orders of
magnitude are typical. 
\begin{figure}[t]
\centering
\includegraphics[scale=.5]{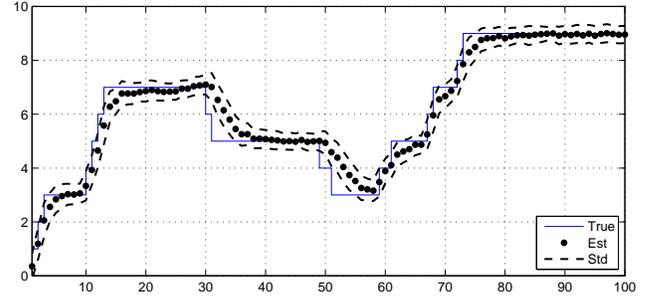}
\caption{Cardinality statistics for Gibbs sampling implementation.}
\label{fig:card_Gibbs}
\end{figure}

\section{Conclusions}

\label{sec:sum} In this paper we propose a new implementation scheme for the 
$\delta$-GLMB filter that allows joint prediction and update. In contrast to
the conventional two-staged implementation, the joint approach use \textit{a
posteriori} information to construct cost matrices for every individual
track, thereby requires only one truncation in each iteration due to the
elimination of inefficient intermediate steps. More importantly, this joint
strategy provides the platform for the development of an accelerated
randomized truncation procedure that achieves superior performance as
compared to that of its traditional deterministic counterpart. The proposed
method is also applicable to approximations of the $\delta$-GLMB filter such
as the labeled multi-Bernoulli (LMB) filter \cite{ReuterLMB14}. 

%

%


\end{document}